\documentclass[nofootinbib,preprint,11pt]{revtex4-1}

\usepackage{tikz}
\usepackage{subfigure}
\usepackage[section]{placeins}
\usepackage{fancyhdr}
\usepackage{array}
\usepackage{mathtools}
\usepackage{amssymb}
\usepackage{graphicx}
\usepackage{epsfig}
\usepackage{wrapfig}
\usepackage{graphics}
\usepackage{slashed}	
\usepackage{enumitem}
\usepackage{color}
\definecolor{SeaBlue}{rgb}{0.1,0.4,0.85}
\definecolor{DarkBlue}{rgb}{0.0,0.0,0.7}
\definecolor{NavyBlue}{rgb}{0.0,0.0,0.4}
\definecolor{Maroon}{rgb}{0.6,0.2,0.2}
\definecolor{SeaGreen}{rgb}{0.2,0.4,0.2}
\definecolor{Purple}{rgb}{0.7,0.1,0.6}
\definecolor{Red}{rgb}{0.8,0.2,0.2}
\definecolor{Black}{rgb}{0.0,0.0,0.0}
\definecolor{ZQS}{rgb}{0.01,0.227,0.451}
\definecolor{TS}{rgb}{0.588,0.31,0.05}
\usepackage[colorlinks]{hyperref}
\hypersetup{
     colorlinks   = true,
     citecolor    = red,
	linkcolor=NavyBlue
}
\usepackage[T1]{fontenc}
\usepackage{titlesec}
\titleformat{\section}
  { \color{NavyBlue} \normalfont \scshape}
  {\thesection}{1em}{\bf \MakeUppercase}
\titleformat{\subsection} { \color{NavyBlue}  \normalfont\scshape}{\thesubsection}{1em}{\bf }{}
\titleformat{\subsubsection}
  { \color{NavyBlue}  \normalfont\itshape}{\thesubsubsection }{1em}{}{}

\usetikzlibrary{decorations.pathmorphing}	
\usetikzlibrary{decorations.markings}
\tikzset{
    vector/.style={decorate, decoration={snake}, draw},
	provector/.style={decorate, decoration={snake,amplitude=2.5pt}, draw},
	antivector/.style={decorate, decoration={snake,amplitude=-2.5pt}, draw},
    fermion/.style={draw=black, postaction={decorate},
        decoration={markings,mark=at position .55 with {\arrow[draw=black]{>}}}},
    fermionbar/.style={draw=black, postaction={decorate},
        decoration={markings,mark=at position .55 with {\arrow[draw=black]{<}}}},
    fermionnoarrow/.style={draw=black},
    gluon/.style={decorate, draw=black,
        decoration={coil,amplitude=4pt, segment length=5pt}},
    scalar/.style={dashed,draw=black, postaction={decorate},
        decoration={markings,mark=at position .55 with {\arrow[draw=black]{>}}}},
    scalarbar/.style={dashed,draw=black, postaction={decorate},
        decoration={markings,mark=at position .55 with {\arrow[draw=black]{<}}}},
    scalarnoarrow/.style={dashed,draw=black},
    electron/.style={draw=black, postaction={decorate},
        decoration={markings,mark=at position .55 with {\arrow[draw=black]{>}}}},
	bigvector/.style={decorate, decoration={snake,amplitude=4pt}, draw},
}

\tikzstyle{block} = [draw, rectangle, 
    minimum height=3em, minimum width=6em]

\definecolor{cerulean}{rgb}{0., 0.52,0.65}

\newcolumntype{P}[1]{>{\raggedright\arraybackslash}p{#1}}

\usepackage{aas_macros}

\makeatletter
\def\l@subsubsection#1#2{}
\makeatother

\begin{document}

\title{\color{NavyBlue} \Large Asymmetric dark matter with a spontaneously broken $U(1)'$: self-interaction and gravitational waves}
\author{\large Zien Chen}
\affiliation{\large Department of Physics and Siyuan Laboratory, Jinan University, Guangzhou 510632, P.R. China }
\author{\large Kairui Ye}
\affiliation{\large Department of Physics and Siyuan Laboratory, Jinan University, Guangzhou 510632, P.R. China }
\author{\large Mengchao Zhang}
\email{Corresponding author: mczhang@jnu.edu.cn}
\affiliation{\large Department of Physics and Siyuan Laboratory, Jinan University, Guangzhou 510632, P.R. China }
\date{\today}
\begin{abstract} 
\normalsize Motivated by the collisionless cold dark matter small scale structure problem, we propose an  asymmetric dark matter model 
where dark matter particle interact with each other via a massive dark gauge boson. 
This model easily avoid the strong limits from cosmic microwave background (CMB) observation, and have a large parameter space to be consistent with small scale structure  data. 
We focus on a special scenario where portals between dark sector and visible sector are too weak to be detected by traditional methods. 
We find that this scenario can increase the effective number of neutrinos ($N_{\text{eff}}$).
In addition, the spontaneous $U(1)'$ symmetry breaking process, which makes dark gauge boson massive, can generate stochastic gravitational waves with peak frequency around $10^{-6} - 10^{-7} \text{ Hz}$.
\end{abstract}
\maketitle
\setcounter{secnumdepth}{3}
\setcounter{tocdepth}{-2}
\tableofcontents

\newpage

\section{Introduction \label{sec:introduction} }

There have been plenty of evidence for the existence of dark matter (DM)~\cite{Ade:2015xua,Clowe:2006eq}, but the nature of dark matter still remains to be revealed. 
Collisionless cold DM is consistent with the large scale structure of the Universe~\cite{Springel:2006vs,Blumenthal:1984bp}. 
However, N-body simulations of collisionless cold DM show some discrepancies between predictions and observations on the scale smaller than $\mathcal{O}$(Mpc).
Those discrepancies include core-cusp problem~\cite{Flores:1994gz,Moore:1994yx,Moore:1999gc}, diversity problem~\cite{Oman:2015xda}, missing satellites problem~\cite{Klypin:1999uc,Moore:1999nt}, and too-big-to-fail (TBTF) problem~\cite{Boylan-Kolchin:2011qkt,Boylan-Kolchin:2011lmk}. 
Including baryon effects in the simulation helps to alleviate some tension~\cite{Navarro:1996bv,Governato:2009bg},
but it is still unclear whether baryon effects can solve all the small-scale problems.  

Small-scale problems may indicate that our assumption about DM, i.e. collisionless, needs to be modified. 
As first pointed out in~\cite{Spergel:1999mh}, a large elastic scattering cross-section between DMs can solve the core-cusp and missing satellites problems. 
Since then, the DM self-interaction was studied by a series of works via N-body simulations~\cite{Rocha:2012jg,Peter:2012jh,Moore:2000fp,Yoshida:2000bx,Burkert:2000di,Kochanek:2000pi,Yoshida:2000uw,Dave:2000ar,Colin:2002nk,Vogelsberger:2012ku,Zavala:2012us,Elbert:2014bma,Vogelsberger:2014pda,Fry:2015rta,Dooley:2016ajo}. 
Recent studies have shown that a velocity independent cross-section between DM is not favored by simulation or semi-analytical results.  
In a nutshell, in dwarf galaxies (where DM velocity $v_{\text{DM}} \sim 10$-$100$ km/s) 
the cross-section per unit mass $\sigma/m_{\text{DM}}$ needs to be in the range $\mathcal{O}(1)$-$\mathcal{O}(10)$ cm$^{2}$/g to solve core-cusp and TBTF problems~\cite{Zavala:2012us,Elbert:2014bma},
 but studies of galaxy groups ($v_{\text{DM}} \sim $1000 km/s) and galaxy clusters ($v_{\text{DM}} \gtrsim$ 1500 km/s) indicate $\sigma/m_{\text{DM}} \sim 0.5 $ cm$^{2}$/g 
and $\sigma/m_{\text{DM}} \lesssim 0.1 $ cm$^{2}$/g respectively~\cite{Harvey:2015hha,Kaplinghat:2015aga,Robertson:2018anx,Sagunski:2020spe,Andrade:2020lqq,Elbert:2016dbb}. 
To be consistent with observables at different scales, a velocity dependent cross-section is required~\cite{Kaplinghat:2015aga}. 
See~\cite{Tulin:2017ara} for a recent review. 

Introducing a light mediator which couples to DM seems to be the easiest way to generate the velocity dependent inter-DM cross-section~\cite{Buckley:2009in,Feng:2009hw,Feng:2009mn,Loeb:2010gj,Kaplinghat:2015aga,vandenAarssen:2012vpm,Tulin:2013teo,Schutz:2014nka,Tulin:2012wi,Boddy:2014qxa,Ko:2014nha,Kang:2015aqa,Kainulainen:2015sva,Wang:2016lvj,Duerr:2018mbd,Kitahara:2016zyb,Ma:2017ucp,Bellazzini:2013foa,Kamada:2020buc,Kamada:2019jch,Kamada:2019gpp,Bringmann:2013vra,Ko:2014bka,Kamada:2018zxi,Kamada:2018kmi,Aboubrahim:2020lnr}. 
Such a scenario is favored in many aspects. 
For example, DM relic density can be realized via the so-called ``secluded freeze-out'' process~\cite{Pospelov:2007mp}, 
which means DM annihilate to light mediators instead of visible SM particles.  
And, because DM relic density and its coupling with SM are unbound, it is easier for such DM models to escape limits from direct detection or collider experiments~\cite{DirectSearch1,DirectSearch2,LHCSearch1,LHCSearch2,LHCSearch3,LHCSearch4,LHCSearch5}. 
However, other studies pointed out that such a "self-interacting DM with light mediator" scenario is strongly constrained by Big Bang nucleosynthesis (BBN), cosmic microwave background (CMB), and indirect search results~\cite{Kamionkowski:2008gj,Zavala:2009mi,Feng:2010zp,Hisano:2011dc,Bergstrom:2008ag,Mardon:2009rc,Galli:2009zc,Slatyer:2009yq,Hannestad:2010zt,Finkbeiner:2010sm}.
This is because the Sommerfeld enhancement induced by the light mediator rapidly increase as DM velocity decreases in the expanding universe~\cite{Sommerfeld,Hisano:2003ec,Hisano:2004ds,Hisano:2005ec,Cirelli:2007xd,Arkani-Hamed:2008hhe,Cholis:2008qq}, 
and thus the energy injection from DM annihilation will affect observables (like BBN or CMB) even after DM freeze-out.  
Especially, the s-wave annihilation case (e.g. DM annihilate to dark gauge boson pair) has been fully excluded by CMB data~\cite{Bringmann:2016din}.  

A simple method to evade those constraints on DM annihilation is to consider the asymmetric dark matter (ADM).
In the ADM scenario, DM is not neutral and self-conjugate, but instead DM is conjugated to anti-DM and the observed DM relic density is determined by the asymmetry between DM and anti-DM. 
See~\cite{Kaplan:2009ag,Petraki:2013wwa,Zurek:2013wia} for recent review. 
When the thermal bath temperature is much lower than the DM mass, the abundance of anti-DM has been reduced to negligible level.
So, the annihilation between DM and anti-DM is much less constrained compared with the symmetric case~\cite{Lin:2011gj,Baldes:2017gzu}. 
Another advantage of ADM model is that it helps to explain the ``$\Omega_{\text{DM}} \simeq 5\Omega_\text{B}$'' coincidence. 
See~\cite{Nussinov:1985xr,Kaplan:1991ah,Barr:1990ca,Barr:1991qn,Dodelson:1991iv,Fujii:2002aj,Kitano:2004sv,Farrar:2005zd,Kitano:2008tk,Gudnason:2006ug,Shelton:2010ta,Davoudiasl:2010am,Huang:2017kzu,Buckley:2010ui,Cohen:2010kn,Frandsen:2011kt,Ibe:2018juk,Ibe:2018tex,An:2009vq,Falkowski:2011xh,Bai:2013xga,Zhang:2021orr,Alves:2009nf,Alves:2010dd,Beylin:2020bsz,Khlopov:1989fj,Blinnikov:1983gh,Blinnikov:1982eh,Blennow:2012de,Murgui:2021eqf,Kamada:2021cow,Ibe:2019ena} for related studies.

In this work we study the DM self-interaction in a concise ADM model framework. 
Combing ADM and DM self-interaction is not a new idea, see e.g.~\cite{Mohapatra:2001sx,Frandsen:2010yj,Petraki:2014uza,Dessert:2018khu,Dutta:2022knf,Heeck:2022znj}.
Compared with previous work, we only consider one flavor of DM (to be labeled as $\chi$) which is charged under a dark $U(1)'$. 
In addition, we introduce two dark Higgs bosons (to be labeled as $S_1$ and $S_2$) charged under the same dark $U(1)'$. 
$S_1$ helps to generate the asymmetry in the dark sector and become dark radiation in the end, and $S_2$ is used to break $U(1)'$ and thus prohibit the long-range interaction between DMs.  
The reason for us to introduce two dark Higgs is that we do not want the troublesome Majorana dark matter mass to be induced by $U(1)'$ symmetry breaking. 
We will clarify this point in the next section. 
To simplify our analysis, we will consider a nearly independent dark sector, which means that the portal between dark sector and visible sector is too small to make these two sectors into thermal equilibrium. 
The portal between two sectors might be too feeble to be searched for via traditional methods like direct detection or collider experiment. 
However, the $U(1)'$ phase transition in the dark sector provides a possible method to detect the dark sector by gravitational waves (GWs), provided the 
phase transition is first order. 
In addition, dark radiation changes the value of effective number of neutrinos ($N_{\text{eff}}$) , which also make this model detectable in the near future.

This paper is organized as following. 
In the next section, we introduce the model framework we want to study. 
In section III we explain how to generate the asymmetry in the dark sector.
We will also discuss the sequential thermal history and related constraints. 
Section III is dedicated to the DM self-interacting and its consistency with data. 
In section IV we discuss the possibility to detect this model via gravitational waves. 
We conclude this work in section V.

\section{Model framework \label{sec:Model} }

In this section we introduce the framework of our model, and specify the scenario we want to study. 

\subsection{Model introduction}

We consider the SM model extended by a dark $U(1)'$ gauged sector.
Similar model framework see~\cite{An:2009vq,Falkowski:2011xh,Dutta:2022knf,Perez:2021udy}.
The Lagrangian can be schematically expressed as: 
\begin{eqnarray}
\mathcal{L} = \mathcal{L}_\text{SM} + \mathcal{L}_\text{Dark} + \mathcal{L}_\text{Portal}
\end{eqnarray} 

$\mathcal{L}_\text{Dark}$ is the Lagrangian of a $U(1)'$ gauged dark sector. 
Dark sector includes a Dirac fermion $\chi$ (dark matter candidate charged under $U(1)'$), dark Higgs $S_1$ (has the same $U(1)'$ charge as $\chi$), and dark Higgs $S_2$ (used to break $U(1)'$ later).
The expression of $\mathcal{L}_\text{Dark}$ is: 
\begin{eqnarray}
\mathcal{L}_\text{Dark} = \bar{\chi}( i \slashed{D} - m_{\chi} ) \chi - (D_{\mu} S_1)^{\dagger} D^{\mu} S_1 - (D_{\mu} S_2)^{\dagger} D^{\mu} S_2 - \frac{1}{4} F'_{\mu\nu}  F'^{\mu\nu} - V(S_1,S_2) 
\end{eqnarray} 
Here $D_{\mu} \equiv \partial_{\mu} + i g' Q_i A'_{\mu} $ ($i=\chi,S_1,S_2$) is the covariant derivative, with $g'$ and $A'_{\mu}$ being the dark gauge coupling and dark gauge boson respectively.
$U(1)'$ charge $\{Q_{\chi}, Q_{S_1}, Q_{S_2}\}$ are simply fixed to $\{+1, +1, +2\}$. 
$F'_{\mu\nu} \equiv \partial_{\mu}A'_{\nu} - \partial_{\nu}A'_{\mu} $ is the field strength of dark gauge boson. 
And $m_{\chi}$ is the mass of dark matter given by hand.  
Dark scalar potential $V(S_1,S_2)$ is:
\begin{eqnarray}
V(S_1,S_2) &=& \mu_1^2 S_1^{\dagger}S_1 - \mu_2^2 S_2^{\dagger}S_2 + \left( \kappa S_2 S_1^{\dagger} S_1^{\dagger} + h.c. \right) \\\nonumber
& &  + \frac{\lambda_1}{4} \left( S_1^{\dagger}S_1 \right)^2 + \frac{\lambda_2}{4} \left( S_2^{\dagger}S_2 \right)^2 + \lambda_{12} \left( S_1^{\dagger}S_1 \right) \left( S_2^{\dagger}S_2 \right)
\end{eqnarray} 
$S_2$ needs to obtain a vacuum expectation value (VEV) after dark phase transition, and thus we insert a minus mass square term $-\mu_2^2$ for it. 
All possible triple and quartic dark Higgs interactions are given. 

$\mathcal{L}_\text{Portal}$ is the sector that connect visible sector and dark sector, including Higgs portal, Abelian gauge boson kinetic mixing, and right handed neutrino (RHN) portal. 
The general expression of $\mathcal{L}_\text{Portal}$ is: 
\begin{eqnarray}
\mathcal{L}_\text{Portal} &=& \frac{1}{2} \sum_{i=1,2} \bar{N}_i ( i \slashed{\partial} - M_{N_i} ) N^C_i - \sum_{i=1,2} \lambda_{S_i H} \left( S_i^{\dagger}S_i \right) \left( H^{\dagger}H \right) - \frac{1}{2} \epsilon F'_{\mu\nu} F^{\mu\nu} \\\nonumber
& & - \sum_{i=1,2} y'_i \bar{N}_i \chi S_1^{\dagger} - \sum_{i=1,2} y_i \bar{N}_i L H^{\dagger} + h.c.
\end{eqnarray} 
Here $\lambda_{S_i H}$ and $\epsilon$ are the coupling of Higgs portal and kinetic mixing parameter respectively. 
Two Majorana RHN, $N_1$ and $N_2$, are introduced to generate the asymmetry in dark or visible sector, with the help of complex phases of $y'_i$ and $y_i$. 
$L$ and $H$ are the SM lepton doublet and Higgs doublet. 

Now we explain the reason to introduce two dark Higgs $S_1$ and $S_2$. 
Assuming that there is no $S_2$ and $U(1)'$ is broken by VEV $\left<S_1\right>$, then, by integrating out $N_1$, a Majorana DM mass ( $\sim \frac{ (y'_1\left<S_1\right> )^2 }{M_{N_1}}$) will be induced. 
This Majorana mass term makes DM oscillate to anti-DM in the late universe.
Thus the asymmetry in the dark sector will be partly erased, and our model will be more limited~\cite{Buckley:2011ye,Tulin:2012re}. 
To make DM stable during the universe lifetime, $M_{N_1}$ needs to be even higher than Planck scale. 
So, to forbid the annoying DM-anti-DM oscillation, in this work we introduce another dark Higgs $S_2$ to break $U(1)'$ and keep $\left<S_1\right> = 0$ all the way.

\subsection{Our scenario}

The model we introduced above is nearly the minimal model that can generate matter asymmetry and induce velocity dependent DM self-interaction.  
However, even for such a nearly minimal model, there are still a dozen parameters to be fixed. 
Diverse and complex phenomena can occur in different parameter spaces, which is difficult to be covered in a single paper. 
Thus, in this paper we choose a simplified scenario to analyze, instead of studying the entire allowed parameter space.

The first simplification we will perform is neglecting $y_i \bar{N}_i L H^{\dagger}$, which is used to generate visible matter asymmetry. 
The inclusion of $y_i \bar{N}_i L H^{\dagger}$ inevitably entangle asymmetries in dark sector and visible sector~\cite{Falkowski:2011xh}, and force us to consider the limits from neutrino data~\cite{Dutta:2022knf}. 
So, in order to focus on phenomena in the dark sector, we are temporarily agnostic to baryon asymmetry problem and neglect $y_i \bar{N}_i L H^{\dagger}$. 

Secondly, we require all the other portals' couplings, i.e. $\lambda_{S_1 H}$, $\lambda_{S_2 H}$, and $\epsilon$, to be small enough to avoid current limits from terrestrial experiments. Furthermore, we also require that these portals are too weak to keep dark sector and visible sector in the thermal equilibrium from reheating to current time. 
These requirements are made for simplicity. 
However, it is also important to study the detectability of this extreme scenario. 
As we will show later, stochastic GWs and the change of $N_{\text{eff}}$ are possible detection methods.


\section{Thermal history of the dark sector and its parameter bounds}

Before the thermal history analysis, in Tab.~\ref{particles} we present all the particles in the dark sector.
Their mass range and the role they played are also given. 
The mass of dark matter ($m_{\chi}$) and dark mediator ($m_{A'}$) are chosen to be consistent with the small scale data. 
To generate asymmetry in the dark sector, the decay of $N_1$ needs to be out-of-equilibrium, and thus the mass of $N_1$ should be much larger than its decay products. 
The mass of $s_2$ ($s_2$ is the scalar component of $S_2$ after $U(1)'$ breaking) is chosen to be smaller than $m_{A'}$. As we will explain later, this is necessary if the $U(1)'$ symmetry breaking is a first order phase transition.  
Finally, the entropy in the dark sector should go to some nearly massless particles long after DM-anti-DM annihilation, otherwise there will be overclosure problem~\cite{Blennow:2012de}. 
So we require $S_1$ to be very light and serve as dark radiation. 

\begin{table}[htp]
\begin{center}
\begin{tabular}{ c c c }
\hline
\hline
name &\ \ \ mass range  \ \ \ &  \ \ \  role  \ \ \  \\ 
\hline
\hline
$\chi$ & \ \ \  10 GeV -- 100 GeV \ \ \  & \ \ \ dark matter \ \ \  \\
\hline
$A'$ & \ \ \  1 MeV -- 100 MeV \ \ \  & \ \ \ mediator between DMs \ \ \  \\ 
\hline
$N_1,N_2$ & \ \ \  $M_{N_i} \gg m_\chi$,\  $M_{N_2} > M_{N_1}$  \ \ \  & \ \ \ generate DM-anti-DM asymmetry \ \ \  \\ 
\hline
$S_2$ & \ \ \  $m_{s_2} < m_{A'}$  \ \ \  & \ \ \ break $U(1)'$ symmetry  \ \ \  \\ 
\hline
$S_1$ & \ \ \  $ m_{S_1} \ll $ 1 eV \ \ \  & \ \ \ dark radiation \ \ \  \\ 
\hline
\end{tabular}
\caption{Particle content in the dark sector, with their mass range and role given.}
\label{particles}
\end{center}
\end{table}


Furthermore, we define the ratio between dark sector temperature $T'$ and visible sector temperature $T$:
\begin{eqnarray}
\xi \equiv \frac{T'}{T}
\end{eqnarray}
The value of $\xi$ will be different in different period. 
In this work we assume the dark sector and visible sector thermally decoupled very early, 
then these two sectors evolve independently. 
The temperature ratio $\xi$ at the time when dark sector temperature $T'$ is lower than $M_{N_2}$ and higher than $M_{N_1}$, is labeled by $\xi_\text{ini}$, and we take it as an input parameter.  

Co-moving entropy densities in each sector are conserved respectively. 
So the temperature ratio in different period will be rescaled by the effective numbers of relativistic degree of freedom (d.o.f.) in each sectors (to be labeled as $g'_{\star}$ and $g_{\star}$)\footnote{Strictly speaking,  $g_{\star}$ for energy density and entropy density are different. 
But before the neutrino decoupling, relativistic d.o.f. for energy density and entropy density in the visible sector are the same. } at that time:
\begin{eqnarray}
\xi = \xi_{\text{ini}} \left( \frac{g_{\star} \cdot g'_{\star,\text{ini}} }{ g'_{\star} \cdot g_{\star,\text{ini}} } \right)^{\frac{1}{3}}
\label{T_ratio}
\end{eqnarray}
In this work we assume $g_{\star,\text{ini}}$ to be the SM value 106.75~\cite{Husdal:2016haj}. 
For the dark sector, $g'_{\star,\text{ini}}$ comes from $N_1$, $\chi$, $S_1$, $S_2$, and $A'$. And so: 
\begin{eqnarray}
g'_{\star,\text{ini}} = \frac{7}{8} \left( 2 + 4 \right) + \left( 2 + 2 + 2 \right) = 11.25
\end{eqnarray}
Given two initial values $g_{\star,\text{ini}}$ and $g'_{\star,\text{ini}}$, temperature ratio $\xi$ at a later time can be determined. 

During the radiation dominant period, energy and entropy densities are given by: 
\begin{eqnarray}
\left\{
          \begin{array}{l}
          \rho = \frac{\pi^2}{30} ( g_{\star}/\xi^4  +  g'_{\star} ) {T'}^4 = \frac{\pi^2}{30} g_{\text{eff}}(T') {T'}^4  \\
          s = \frac{2 \pi^2}{45} ( g_{\star}/\xi^3  +  g'_{\star} ) {T'}^3 = \frac{2 \pi^2}{45} h_{\text{eff}}(T') {T'}^3
          \end{array}
\right.
\end{eqnarray}
Here we define the effective d.o.f. for energy and entropy, $g_{\text{eff}}(T')$ and $h_{\text{eff}}(T')$, for later convenience.

\subsection{The generation of dark sector asymmetry}

\begin{figure}[ht]
\centering
\includegraphics[width=6.0in]{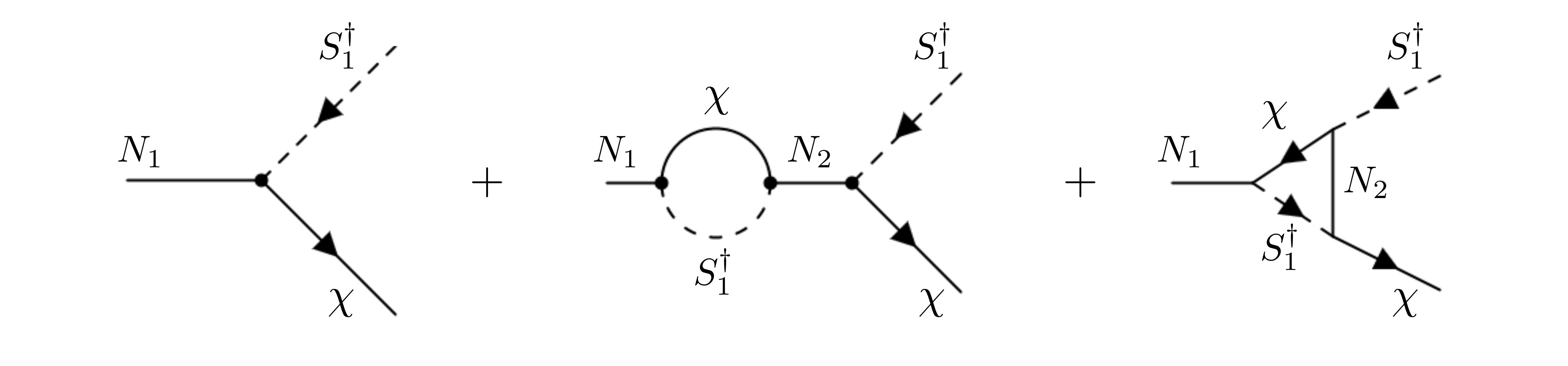}
\caption{CP violated process in the dark sector that generates the asymmetry between DM and anti-DM.}
\label{CPV}
\end{figure}

In this subsection we introduce the generation of $Y_{\Delta \chi} \equiv Y_{\chi} - Y_{\bar{\chi}}$.
Here $Y$ is the particle yield which equals to particle number density divided by entropy density. 
Before the $U(1)'$ symmetry breaking, $U(1)'$ charge is conserved and thus $Y_{\Delta S_1} \equiv Y_{S_1} - Y_{S_1^{\dagger}} = - Y_{\Delta \chi}$. 
Similar to the vanilla leptogenesis~\cite{Fukugita:1986hr,Buchmuller:2004nz,Davidson:2008bu,Covi:1996wh}, non-zero $Y_{\Delta \chi}$ is generated by the CP violated and out-of-equilibrium decay of $N_1$. 
See Fig.(\ref{CPV}) for illustration. 

Asymmetric yield $Y_{\Delta \chi}$ can be expressed as:
\begin{eqnarray}
Y_{\Delta \chi} = Y_{N_1}\times \epsilon \times \eta
\end{eqnarray}
Here $Y_{N_1}$ is the yield of $N_1$ before it decays. 
Because $N_1$ is in the equilibrium with dark thermal bath initially, so the initial yield of $N_1$ is: 
\begin{eqnarray}
Y_{N_1} = \left( \frac{3}{4} 2 \frac{\zeta(3)}{\pi^2} T'^3 \right) \div  \left[ \frac{2\pi^2}{45} \left( g'_{\star,\text{ini}} T'^3 +  g_{\star,\text{ini}} T^3  \right) \right]
\simeq \frac{0.42}{11.25 + 106.75 / \xi_{\text{ini}}^{3} }
\end{eqnarray}

$\epsilon$ is the CP asymmetry generated by $N_1$ decay: 
\begin{eqnarray}
\epsilon \equiv \frac{\Gamma(N_1 \to \chi S_1^{\dagger}) - \Gamma(N_1 \to \bar{\chi} S_1)}{\Gamma(N_1 \to \chi S_1^{\dagger}) + \Gamma(N_1 \to \bar{\chi} S_1)}
\end{eqnarray}
The expression of $\epsilon$ can be simplified when $M_{N_2} \gg M_{N_1}$. In this case, $\epsilon$ is approximately given by: 
\begin{eqnarray}
\epsilon \simeq -\frac{1}{16\pi} \frac{M_{N_1}}{M_{N_2}} \frac{ \text{Im} \left[ (y'^{\ast}_2 y'_1 )^2 \right] }{|y'_1|^2}
\end{eqnarray}

\begin{figure}[ht]
\centering
\includegraphics[width=4.5in]{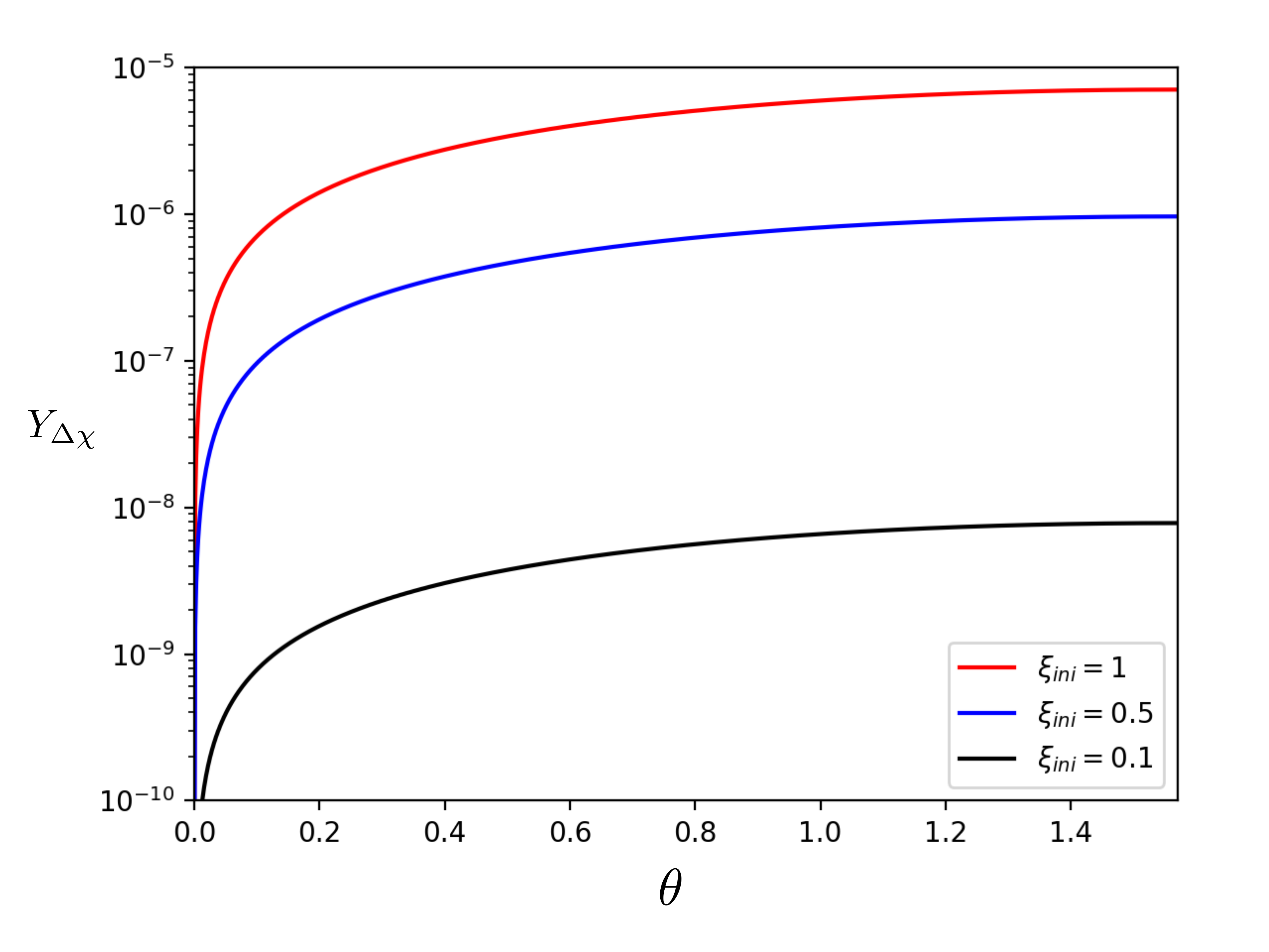}
\caption{$Y_{\Delta \chi} \equiv Y_{\chi} - Y_{\bar{\chi}}$ as functions of CP phase angle $\theta$ with different $\xi_{\text{ini}}$ values.}
\label{Y_chi}
\end{figure}

$\eta$ is the efficiency factor that reduce the final generated asymmetry. 
In the so-called ``weak washout'' case where the decay width of $N_1$ is smaller than the Hubble expansion rate ($\Gamma_{N_1} < H(T' = M_{N_1}) $), 
the value of $\eta$ can be close to 1. 
To simplify our analysis we will only consider ``weak washout'' case, and it leads to a constraint on the parameter space: 
\begin{eqnarray}
\Gamma_{N_1} & < & H(T' = M_{N_1}) \\
\Rightarrow \ \ \ \  \frac{|y'_1|^2}{16\pi} M_{N_1}  & < &  1.66 \frac{M^2_{N_1}}{ M_{\text{Pl}} } \sqrt{  g'_{\star,\text{ini}} +  g_{\star,\text{ini}} / \xi_{\text{ini}}^{4}  } \\
\Rightarrow \ \ \ \   M_{N_1}  & > &  \frac{ |y'_1|^2 M_{\text{Pl}} }{ 83.44 \times \sqrt{  11.25 + 106.75 / \xi_{\text{ini}}^{4}  }  }
\end{eqnarray}
Here $M_{\text{Pl}} \simeq 1.22 \times 10^{19}$ GeV is the Planck mass. 
So there is a large parameter space to satisfy the ``weak washout'' requirement.

For convenience, we define CP phase angle $\theta$ by:
\begin{eqnarray}
(y'^{\ast}_2 y'_1 )^2 = |y'_2|^2 |y'_1|^2 e^{i\theta}
\end{eqnarray}
In Fig. (\ref{Y_chi}) we show $Y_{\Delta\chi}$ as functions of CP phase angle $\theta$ with $\frac{M_{N_1}}{M_{N_2}}=0.1$ and $\xi_{\text{ini}}$ fixed to 1, 0.5, and 0.1 respectively.
It can be seen that even for very small $\xi_{\text{ini}}$, $Y_{\Delta\chi}$ can exceed $10^{-9}$.
DM relic density can be estimated by: 
\begin{eqnarray}
\Omega_{\chi} h^2 \approx m_{\chi} Y_{\Delta\chi} s_0 / \rho_{\text{cr}} \approx Y_{\Delta\chi} \left( \frac{m_{\chi}}{\text{GeV}} \right) \times 2.72 \times 10^8
\end{eqnarray}
For dark matter mass $m_{\chi}$ larger than 10 GeV, it is always possible to explain current observed relic density ($\Omega_{\chi} h^2 \approx 0.12$), provided $\xi_{\text{ini}}$ is not much smaller than 1. 
So we can take $Y_{\Delta\chi}$ as an input parameter, which should be consistent with $\Omega_{\chi} h^2 \approx 0.12$, in the following analysis.

\subsection{$\chi - \bar{\chi}$ annihilation \label{sec:ann} }


As we explained in the introduction, asymmetry DM helps to escape the limits from observations like CMB. 
To be more specific, compared with symmetric DM scenario, the energy injection rate in ADM scenario during recombination is suppressed by the asymptotic ratio: 
\begin{eqnarray}
r_{{\infty}} =  \frac{Y_{\bar{\chi}}(\infty)}{Y_{{\chi}}(\infty)} 
\end{eqnarray}
To obtain $r_{{\infty}}$ (here "$\infty$" correspond to recombination time), we need to solve following Boltzmann equations for yields $Y_{\chi}$ and $Y_{\bar{\chi}}$~\cite{Scherrer:1985zt,Griest:1986yu,Graesser:2011wi,Iminniyaz:2011yp,Bell:2014xta,Murase:2016nwx,Baldes:2017gzw}:
\begin{eqnarray}
\frac{dY_{\chi,\bar{\chi}}}{dx} = -  \frac{m_{\chi} M_{\text{Pl}} }{x^2}  \sqrt{\frac{\pi g_\ast }{45}} \left< \sigma_{\text{ann}} v \right> \left(Y_{\chi}Y_{\bar{\chi}} -  Y^{\text{sym}}_{\text{eq}} Y^{\text{sym}}_{\text{eq}}   \right)
\label{BZE}
\end{eqnarray}
Here $x \equiv m_{\chi}/T' $, and $Y^{\text{sym}}_{\text{eq}}$ is the equilibrium yield of $\chi$ (or $\bar{\chi}$) with chemical potential being zero (correspond to the symmetric case):  
\begin{eqnarray}
Y^{\text{sym}}_{\text{eq}} = \left( \frac{2}{(2\pi)^3} \int \frac{d^3 \vec{p}}{e^{\sqrt{m^2_{\chi}+\vec{p}^2}/T'}+1} \right) \div  \left[  \frac{2 \pi^2}{45} h_{\text{eff}}(T') {T'}^3 \right]
\end{eqnarray}
And:
\begin{eqnarray}
\sqrt{g_\ast} =  \frac{ h_{\text{eff}} }{\sqrt{ g_{\text{eff}}  }} \left(1+ \frac{T'}{3 h_{\text{eff}}} \frac{d h_{\text{eff}}}{d T'}  \right) 
\end{eqnarray}


Ratio $r(x)$ is a function of $x$, and the asymptotic ratio $r_{{\infty}}$ is the value of $r(x)$ when $x\to\infty$:
\begin{eqnarray}
r(x) = \frac{Y_{\bar{\chi}}(x)}{Y_{{\chi}}(x)} \ , \  r_{{\infty}} = \lim_{x\to \infty} r(x)
\end{eqnarray}
We follow the method proposed in~\cite{Baldes:2017gzw,Graesser:2011wi} to calculate $r_{{\infty}}$. 
 For later convenience, firstly we need to define equilibrium ratio $r_{\text{eq}}(x)$ as:
\begin{eqnarray}
r_{\text{eq}}(x) = \frac{Y^{\text{sym}}_{\text{eq}} \times \exp\left[ - \sinh^{-1} \left( \frac{Y_{\Delta\chi}}{2 Y^{\text{sym}}_{\text{eq}}}    \right) \right]   }{Y^{\text{sym}}_{\text{eq}} \times \exp\left[ + \sinh^{-1} \left( \frac{Y_{\Delta\chi}}{2 Y^{\text{sym}}_{\text{eq}}}    \right) \right] } = \exp\left[ -2 \sinh^{-1} \left( \frac{Y_{\Delta\chi}}{2 Y^{\text{sym}}_{\text{eq}}}    \right) \right]
\end{eqnarray}
Here, ``$\sinh^{-1} \left( \frac{Y_{\Delta\chi}}{2 Y^{\text{sym}}_{\text{eq}}}    \right)$'' is actually the ratio between chemical potential and temperature. 
Then Boltzmann equations~(\ref{BZE}) can be transferred to a differential equation for $r(x)$~\cite{Graesser:2011wi}: 
\begin{eqnarray}
\frac{d r }{dx} = -   \frac{m_{\chi} M_{\text{Pl}} }{x^2}  \sqrt{\frac{\pi g_\ast }{45}} \left< \sigma_{\text{ann}} v \right>  Y_{\Delta\chi} 
 \left[   r - r_{\text{eq}}\left( \frac{1-r}{1-r_{\text{eq}} } \right)^2   \right]
\label{ratio1}
\end{eqnarray}

Before freeze-out ($x<x_{FO}$), $\chi$ and $\bar{\chi}$ are in the thermal equilibrium and thus $r=r_{\text{eq}}$. 
After freeze-out ($x>x_{FO}$), $r_{\text{eq}}$ decreases much faster than $r$ and thus the Eq.~(\ref{ratio1}) can be approximatively simplified to: 
\begin{eqnarray}
\frac{d r }{dx} \simeq -   \frac{m_{\chi} M_{\text{Pl}} }{x^2}  \sqrt{\frac{\pi g_\ast }{45}} \left< \sigma_{\text{ann}} v \right>  Y_{\Delta\chi} \ r
\label{ratio2}
\end{eqnarray}
Then we obtain the approximate expression of $r_{{\infty}}$:
\begin{eqnarray}
r_{{\infty}} \simeq r(x_{FO}) \exp\left[-m_{\chi}M_{\text{Pl}} Y_{\Delta\chi} \int_{x_{FO}}^{\infty} \sqrt{\frac{\pi g_\ast }{45}}  \frac{\left< \sigma_{\text{ann}} v \right>}{x^2} dx  \right]
\label{ratio_exp1}
\end{eqnarray}
At the freeze-out temperature ($x=x_{FO}$), these is little difference between $r$ and $r_{\text{eq}}$. So Eq.~(\ref{ratio_exp1}) can be further simplified to: 
\begin{eqnarray}
r_{{\infty}} \simeq r_{\text{eq}}(x_{FO}) \exp\left[-m_{\chi}M_{\text{Pl}} Y_{\Delta\chi} \int_{x_{FO}}^{\infty} \sqrt{\frac{\pi g_\ast }{45}}  \frac{\left< \sigma_{\text{ann}} v \right>}{x^2} dx  \right]
\label{ratio_exp2}
\end{eqnarray}

In this work, we consider dark matter within mass range  10 GeV -- 100 GeV. 
Previous numerical study~\cite{Baldes:2017gzw} shows that, within this mass range, the inclusion of non-perturbative effects (i.e. Sommerfeld enhancement and bound state formation) is not important in the calculation of $r_{\infty}$~\footnote{However, for dark matter heavier than TeV, non-perturbative effects play key role in $r_{\infty}$ estimation. See Rf.~\cite{Baldes:2017gzw} for more details.}. 
Thus we can approximately replace cross-section by its leading-order perturbative value: 
\begin{eqnarray}
 (\sigma_{\text{ann}} v) \simeq (\sigma_{\text{ann}} v)_0 = \frac{\pi \alpha'^2}{m^2_{\chi}} Q_\chi^4 
\end{eqnarray}
Here $\alpha' \equiv q'^2/4\pi$ is the dark fine structure constant. 
And $Q_\chi$ has been fixed to 1 in this work. By this approximation, Eq.~(\ref{ratio_exp2}) is further simplified to: 
\begin{eqnarray}
r_{{\infty}} \simeq r_{\text{eq}}(x_{FO}) \exp\left[- \frac{M_{\text{Pl}} Y_{\Delta\chi}}{m_{\chi}} \sqrt{\frac{\pi^3  }{45}} \alpha'^2
\int_{x_{FO}}^{\infty} \frac{\sqrt{g_\ast}}{x^2} dx  \right]
\label{ratio_exp3}
\end{eqnarray}

Finally, freeze-out temperature is determined by: 
\begin{eqnarray}
x_{FO} - \frac{1}{2} \ln x_{FO} \simeq \ln \left[ 0.076\times m_{\chi} M_{\text{Pl}}\ (\sigma_{\text{ann}} v)_0\ \frac{\sqrt{g^{FO}_\ast}}{h^{FO}_{\text{eff}}}    \right]
\label{FO}
\end{eqnarray}
With all the above information, we can estimate $r_{\infty}$ numerically.

\begin{figure}[ht]
\centering
\includegraphics[width=5.5in]{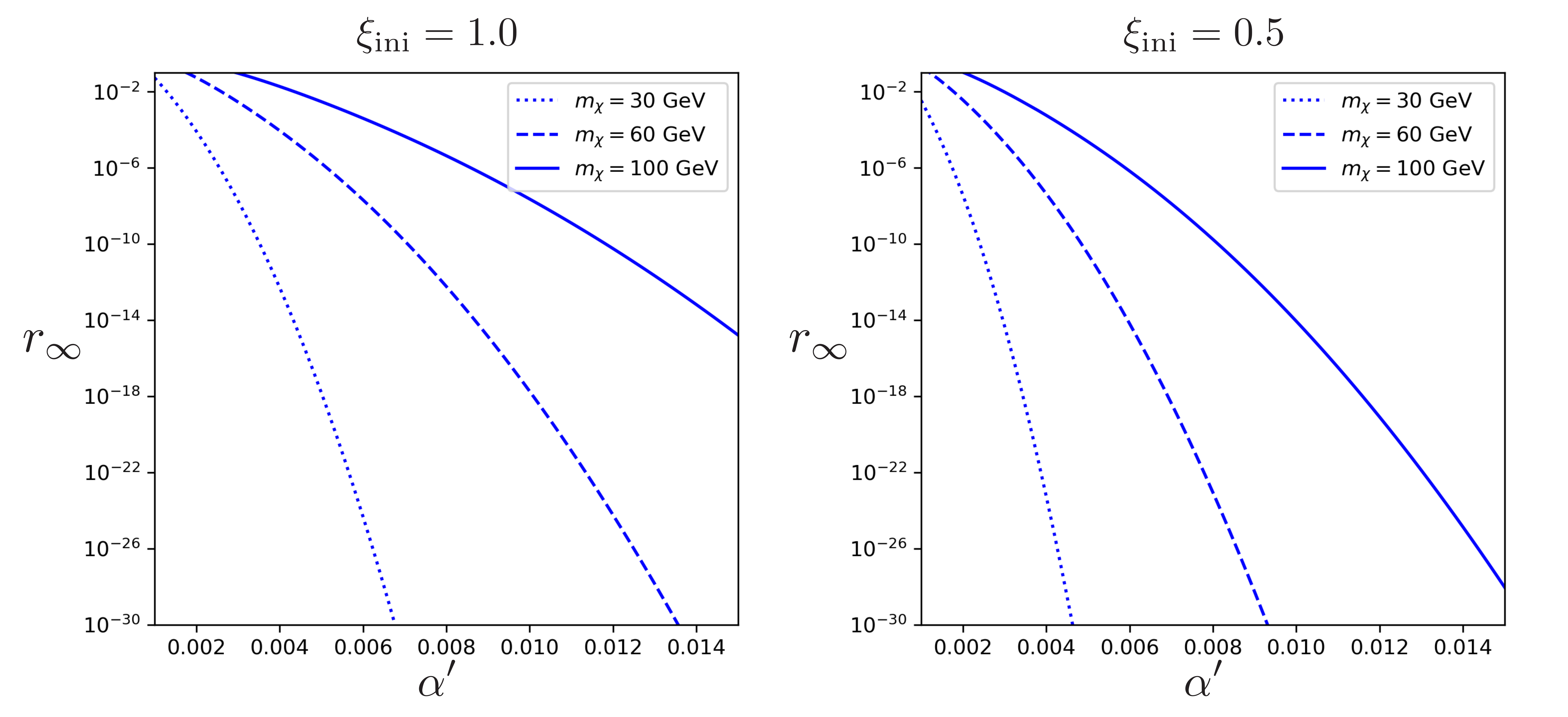}
\caption{ $r_{\infty}$ as functions of dark fine structure constant $\alpha'$, with different dark matter mass $m_{\chi}$ and different initial temperature ratio $\xi_{\text{ini}}$. }
\label{r_infty}
\end{figure}

In Fig.~(\ref{r_infty}) we present the value of $r_{\infty}$ as functions of $\alpha'$. 
It clearly shows that $r_{\infty}$ is very sensitive to the value of $\alpha'$. 
With $\alpha'$ increasing from $\mathcal{O}$(0.001) to $\mathcal{O}$(0.01), $r_{\infty}$ decreases by more than 10 orders. 
The decreasing of $r_{\infty}$ decreasing becomes much quick for smaller dark matter mass. 
This trend is consistent with previous study~\cite{Graesser:2011wi}. 
We also present the dependence of $r_{\infty}$ on temperature ratio, and our results show that $r_{\infty}$ will be smaller if dark sector is colder than the visible sector. 
This relationship can be understood by the enhanced $\sqrt{g_{\ast}}$ during freeze-out when dark sector become colder. 

\subsection{Limit on $\chi-\bar{\chi}$ annihilation during recombination \label{sec:CMB} }

As we explained in the introduction, the strong limit of CMB date on dark matter annihilation during recombination period can be greatly weakened by the asymmetry of dark matter.  
However, due to the scenario we chosen to study in this work, this problem is ``over-solved''. 
In our scenario, we let $S_1$ to be nearly massless and serve as dark radiation. 
So the dominant decay channel of mediator is $\gamma' \to S_1^{\dagger} S_1 $, and the energy injection from $\chi-\bar{\chi}$ annihilation goes to dark thermal bath instead of visible sector. 
Thus the already bonded neutral hydrogen atoms will not be reionized by high energy electric shower process, and $\chi-\bar{\chi}$ annihilation in our scenario is save from the direct CMB limit.

But is still very interesting to see how the asymmetry helps to weaken the CMS limit. 
So in this subsection we will deviate from our scenario and assume that the mediator dominantly decay to electron. 
In this case, BBN might give a strong bound on the mass and lifetime on $\mathcal{O}(1) - \mathcal{O}(10) $ MeV mediator (see e.g. Ref.~\cite{Hufnagel:2018bjp,Depta:2020zbh,Ibe:2021fed} for detailed discussion). 
But here we will only focus on the CMB bound.

As we said in the last subsection, non-perturbative effects in annihilation process can be ignored in the calculation of $r_{\infty}$ for dark matter lighter than 100 GeV. 
But in the study of energy injection during recombination, including the non-perturbative effects in  annihilation is important. 
Here we perform an approximate analysis like~\cite{Bringmann:2016din}, which only include the Sommerfeld enhancement in the estimation of annihilation cross-section during recombination period. 

The annihilation cross-section can be written as the tree-level cross-section multiplied by a Sommerfeld enhancement factor~\cite{Cassel:2009wt}: 
\begin{eqnarray}
 (\sigma_{\text{ann}} v) = S(v) \times (\sigma_{\text{ann}} v)_0 
\end{eqnarray}
Tree level annihilation cross section $(\sigma_{\text{ann}} v)_0  = {\pi \alpha'^2Q_\chi^4}/{m^2_{\chi}} $ have been given in previous subsection. 
Sommerfeld enhancement factor $S(v)$ is:
\begin{eqnarray}
S(v) = \frac{\pi}{\mathcal{A}(v)} \frac{\sinh(2\pi \mathcal{A}(v) \mathcal{B} )}{\cosh(2\pi \mathcal{A}(v) \mathcal{B}) - 
\cos(2\pi \sqrt{ \mathcal{B} - (\mathcal{A}(v) \mathcal{B})^2}) }  
\end{eqnarray}
with: 
\begin{eqnarray}
\mathcal{A}(v) = \frac{v}{2 Q^2_{\chi} \alpha'}  \ , \
\mathcal{B} = \frac{6Q_{\chi}^2 \alpha' m_{\chi}}{\pi^2 m_{\gamma'}}
\end{eqnarray}

Sommerfeld enhancement factor will reach it maximal value, or say saturates, when velocity $v \lesssim m_{\gamma'}/2m_{\chi} $.
During recombination period, this saturation condition is already satisfied~\cite{Bringmann:2016din}, and thus the annihilation cross-section will generally be enhanced by several orders. 

\begin{figure}[ht]
\centering
\includegraphics[width=6.5in]{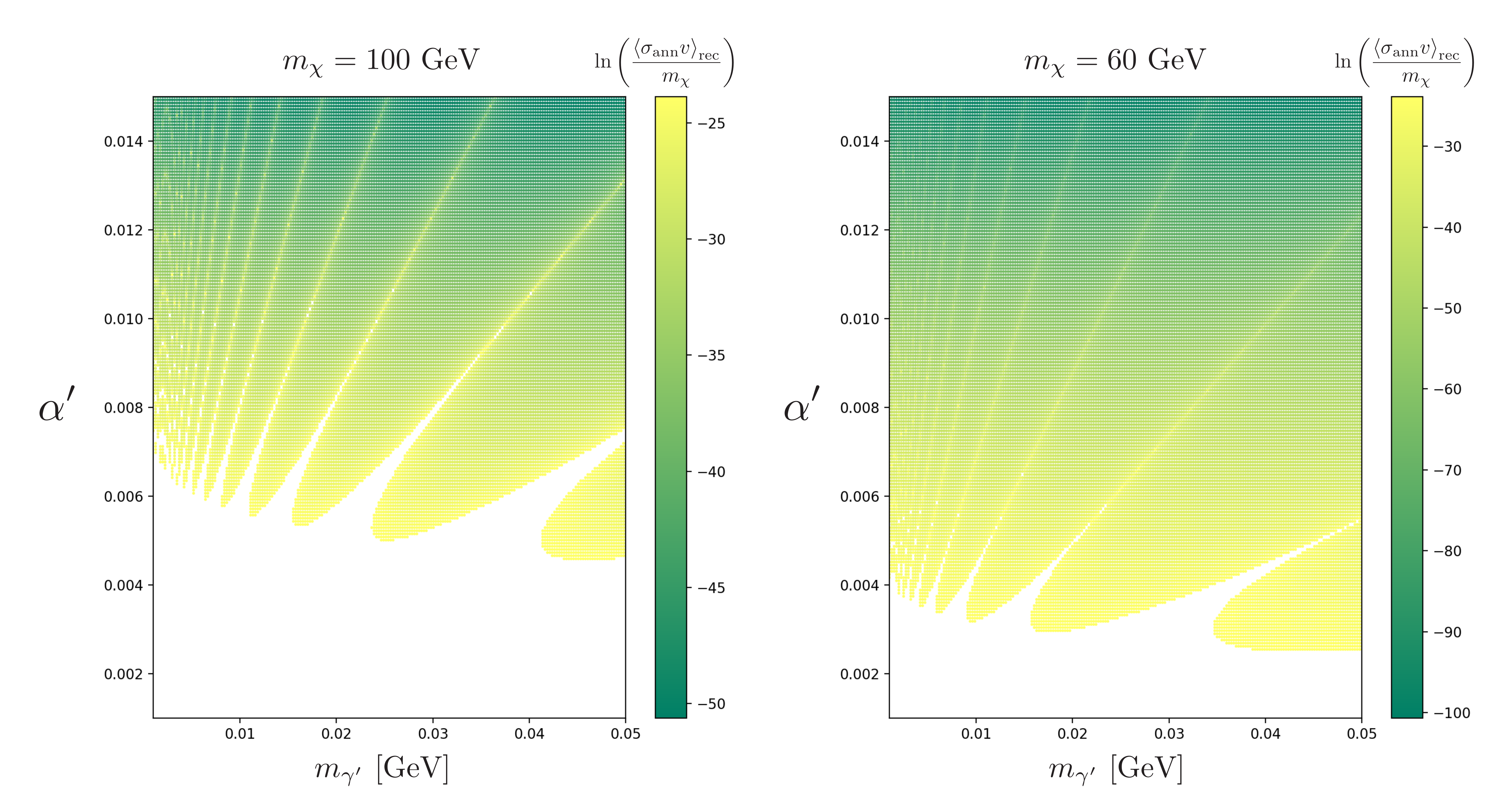}
\caption{ CMB limit on $\alpha'$-$m_{\gamma'}$ plane with $m_{\chi}$ fixed to different values. 
In this plot we assume that the $\chi - \bar{\chi}$ annihilation mainly go to electron final states.  
The color indicate the natural logarithmic value of $\frac{\left< \sigma_{\text{ann}} v \right>_{\text{rec}}}{m_\chi}$ in units of $\text{GeV}^{-3}$. 
The point with $\ln \left( \frac{\left< \sigma_{\text{ann}} v \right>_{\text{rec}}}{m_\chi} \right) >  -23.88 $ (blank region) have been excluded by current data. }
\label{CMB}
\end{figure}

Many studies has been done on the CMB's constraints on dark matter annihilation~\cite{Galli:2009zc,Slatyer:2009yq,Cline:2013fm,Liu:2016cnk}.
Recently study~\cite{Kawasaki:2021etm} proposes a slightly stronger constraint by using combined date from Planck~\cite{Planck:2019nip}, BAO~\cite{BOSS:2013rlg,Ross:2014qpa}, and DES~\cite{DES:2017myr}. 
For electron final states and DM mass within 10 GeV to 100 GeV, limit on $\left< \sigma_{\text{ann}} v \right>_{\text{rec}} /m_\chi $ is (for symmetric DM): 
\begin{eqnarray}
\frac{\left< \sigma_{\text{ann}} v \right>_{\text{rec}}}{m_\chi} \lesssim 5\times 10^{-28} \ \text{cm}^3\text{s}^{-1}\text{GeV}^{-1}\  (  4.26\times 10^{-11}\ \text{GeV}^{-3} )
\label{limit1}
\end{eqnarray}
Here we also given the limit in natural units. 
To illustrate how the CMB limit is, we can consider $\alpha'=0.01$ and $m_{\chi} = 100$ GeV. 
In this case, $\frac{\left< \sigma_{\text{ann}} v \right>_{\text{rec}}}{m_\chi} \sim S(v_{\text{rec}}) \times 10^{-10} \ \text{GeV}^{-3}$. 
So even for $\mathcal{O}(1)$ enhancement factor $S(v_{\text{rec}})$, this parameter choice can not avoid CMB constraint. 
In the next section we will show that $\alpha' \sim \mathcal{O}(0.1)$ is generally required to solve the small scale problem.  
Thus it is very difficult for symmetric DM to be consistent with CMB data, provided the final state of DM annihilation is elections.  

Different with the symmetric DM case, in our asymmetric DM case, this limit should be modified to: 
\begin{eqnarray}
2 r_{\infty} \times \frac{\left< \sigma_{\text{ann}} v \right>_{\text{rec}}}{m_\chi} \lesssim 5\times 10^{-28} \ \text{cm}^3\text{s}^{-1}\text{GeV}^{-1}\  (  4.26\times 10^{-11}\ \text{GeV}^{-3} )
\label{limit2}
\end{eqnarray}
As we mentioned before, the energy injection from DM annihilation during recombination is reduced hugely by the small value of $r_{\infty}$. 
And thus the constrain from CMB to dark sector parameters become much looser.  
In Fig.~\ref{CMB} we present the allowed parameter region with $m_{\chi}$ fixed to 100 GeV and 60 GeV, respectively. 
As we already shown in the last subsection, increasing $\alpha'$ value lead to $r_{\infty}$ exponential decreasing, and thus larger $\alpha'$ is more easier to escape from CMB constrain. 
And for asymmetric DM mass within 10 GeV to 100 GeV, $\alpha' \gtrsim 0.01$ is large enough to escape CMB limit (even if the final state of DM-anti-DM annihilation are electrons). 
This is also favored by small scale data. 

After the discussion of CMB constraint, we move back to our scenario with dark radiation. 

\subsection{Change of $N_{\text{eff}}$  \label{sec:neff} }

As we explain before, in our scenario all the entropy in dark sector finally goes to nearly massless complex scalar $S_2$, which is dark radiation. 
The presence of dark radiation will affect the measured value of the effective number of neutrino species $N_{\text{eff}}$~\cite{Blennow:2012de}.
$N_{\text{eff}}$ is defined by the measured radiative energy density in addition to photon energy density: 
\begin{eqnarray}
\rho_r  = \rho_{\gamma} \left[ 1 + \frac{7}{8} \left( \frac{T_{\nu}}{T_{\gamma}} \right)^4 N_{\text{eff}}     \right]
\label{Neff1}
\end{eqnarray}
Current constraint on $N_{\text{eff}}$ from joint Planck + BAO data analysis is~\cite{Planck:2018vyg}:
\begin{eqnarray}
N_{\text{eff}} = 2.99 ^{+0.34} _{-0.33}  \ \  (95 \%)
\label{Neff2}
\end{eqnarray}
On the other hand the SM prediction of $N_{\text{eff}}$ is~\cite{deSalas:2016ztq}: 
\begin{eqnarray}
N^{\text{SM}}_{\text{eff}} = 3.045
\label{Neff3}
\end{eqnarray}
Thus there is a room about $ \Delta N_{\text{eff}} < 0.29$ for the existence of dark radiation (DR). 

In this work we consider an independent dark sector, and hence ${T_{\nu}}/{T_{\gamma}}$ retain its SM value $(4/11)^{1/3}$. 
Then $\Delta N_{\text{eff}}$ can be expressed as: 
\begin{eqnarray}
\Delta N_{\text{eff}} &=& \frac{8}{7} \left( \frac{ T_{\nu} }{ T_{\gamma} } \right)^{-4} \frac{ \rho_{\text{DR}} }{\rho_\gamma}
 = \frac{8}{7} \left( \frac{ 4 }{ 11 } \right)^{-4/3}  \left( \frac{ T' }{ T_{\gamma} } \right)^{4} \\\nonumber
 &=& \frac{8}{7} \left( \frac{ 4 }{ 11 } \right)^{-4/3} \left( \frac{ 11.25\times 3.91 }{106.75\times 2} \right)^{4/3} \xi_{\text{ini}}^4
\label{Neff4}
\end{eqnarray}
The temperature ratio $T'/T_{\gamma}$ after the second equal sign should be estimated during recombination period by Eq.~(\ref{T_ratio}).
Thus the limit on $ \Delta N_{\text{eff}} $ is transferred to the limit on $\xi_{\text{ini}}$:
\begin{eqnarray}
\xi_{\text{ini}} < 0.86
\label{Neff5}
\end{eqnarray}

It should be noted that this up-limit on $\xi_{\text{ini}}$ needs to be modified when the intensity of dark $U(1)'$ phase transition is large~\cite{Bai:2021ibt}. 
We will discuss this point in the gravitational wave section. 

The future CMB-S4 experiment will constrain the deviation from SM to $\Delta N_{\text{eff}} < 0.06 $ at 95\% C.L.~\cite{CMB-S4:2022ght}.
If the initial temperature ratio $\xi_{\text{ini}}$ is not too small, then we should observe an exceed of $\Delta N_{\text{eff}}$ at CMB-S4.

\subsection{ Dark acoustic oscillations and collisional damping \label{sec:sca} }

The presence of dark radiation (DR) cause another problem which might make our scenario constrained by current cosmology observations. 
In our scenario, DM $\chi$ and DR $S_2$ are both charged under the $U(1)'$, and thus they can scatter with each other via the dark mediator $\gamma'$. 
This DM-DR scattering may cause the so-called ``dark acoustic oscillations'' (DAO) and the collisional (Silk) damping between DM and DR~\cite{Cyr-Racine:2013fsa,Buckley:2014hja}, provided the kinetic equilibrium between DM and DR lasts long enough.
DAO and the collisional (Silk) damping will modify the initial matter power spectrum, and then leave imprints on CMB anisotropy and large scale structure (LSS)~\cite{Cyr-Racine:2013fsa,Buckley:2014hja}. 

Ref.~\cite{Cyr-Racine:2013fsa} propose a parameter $\Sigma_{\text{DAO}}$ as the proxy of DAO effect. 
$\Sigma_{\text{DAO}}$ is related to the scattering cross-section between DM and DR (labeled as $\sigma_{\text{DM-DR}}$) via: 
\begin{eqnarray}
\frac{ \sigma_{\text{DM-DR}}(T'_{\text{dec}}) }{m_{\chi}} = 1.9\times 10^{-4} \left( \frac{\xi_{\text{dec}}}{0.5} \right) \left( \frac{\Sigma_{\text{DAO}}}{10^{-3}} \right) \frac{\text{cm}^2}{\text{g}}
\label{DAO}
\end{eqnarray}
where $T'_{\text{dec}}$ is the DM kinetic decoupling temperature.
 $\sigma_{\text{DM-DR}}(T'_{\text{dec}})$ and $\xi_{\text{dec}}$ are scattering cross-section and temperature ratio at $T'_{\text{dec}}$, respectively. 
The bound on the value of $\Sigma_{\text{DAO}}$ is $\Sigma_{\text{DAO}} < 10^{-4.15}\ (10^{-3.6})$ for $\xi = 0.5$ (0.3)~\cite{Cyr-Racine:2013fsa}.

The kinetic decoupling temperature $T'_{\text{dec}}$ is determined by: 
\begin{eqnarray}
n_{\gamma'} \langle \sigma v \rangle_{ \text{DM-DR} } v^2_{\text{DM}}  \approx H(T'_{\text{dec}})
\end{eqnarray}
The left hand side of the above equation can be approximated by: 
\begin{eqnarray}
n_{\gamma'} \langle \sigma v \rangle_{ \text{DM-DR} } v^2_{\text{DM}}  \approx
 \frac{2.4}{\pi^2} (T'_{\text{dec}})^3 \times \pi \frac{ \alpha'^2 (T'_{\text{dec}})^2 }{ m^4_{\gamma'} } \times \frac{T'_{\text{dec}}}{m_{\chi}}
\end{eqnarray}
Generally speaking, $T'_{\text{dec}}$ is much smaller than $m_{\gamma'}$, so here we estimate $\langle \sigma v \rangle_{ \text{DM-DR} }$ by a simple dimensional analysis. 
On the other hand, $H(T'_{\text{dec}}) \approx 1.66 \frac{ (T'_{\text{dec}})^2 }{ M_{\text{Pl}} } \sqrt{ 2 + 3.4/\xi_{\text{dec}}^4 }$. 
Combined with Eq.~(\ref{DAO}) we can induce a bound on coupling strength and spectrum (here we choose $\xi_{\text{dec}}=0.5$):
\begin{eqnarray}
 \frac{\alpha'}{ m^2_{\gamma'} \sqrt{m_{\chi} M_{\text{Pl}} }  } \lesssim 4.7\times 10^{-3}\ \text{GeV}^{-3}
\end{eqnarray}
So it can be seen that even for $ m_{\chi} =10 $ GeV and $m_{\gamma'} = 1$ MeV, the bound on $\alpha'$ is still very loose.

\section{DM self-interaction and small scale structure \label{sec:SI} }

In this section we investigate under which parameter settings the elastic scattering cross-section between DMs can be consistent with small-scale observations. 
Only $\{ m_{\chi},\ m_{\gamma'},\ \alpha'  \}$ are relevant parameters in this section. 

The calculation methods of DM scattering cross-section depend on the value of $\{ m_{\chi},\ m_{\gamma'},\ \alpha'  \}$ and the relative velocity between DMs. 
Basically, there are four different regimes. 
In the Born regime ( $ \frac{ \alpha'  m_{\chi} }{m_{\gamma'}} \ll 1 $ ), one can do perturbative calculation and obtain analytic formula directly~\cite{Tulin:2013teo,Feng:2009hw,Buckley:2009in,Kahlhoefer:2017umn}. 
In the classical regime ( $ \frac{ \alpha'  m_{\chi} }{m_{\gamma'}} \gtrsim 1 $ and $ \frac{ m_{\chi} v_{\text{rel}} }{ m_{\gamma'} } \gg 1 $ ), numerical results can be fitted with analytical functions~\cite{Feng:2009hw,Buckley:2009in,Khrapak:2003kjw,Cyr-Racine:2015ihg}. 
In the quantum regime (  $ \frac{ \alpha'  m_{\chi} }{m_{\gamma'}} \gtrsim 1 $ and $ \frac{ m_{\chi} v_{\text{rel}} }{ m_{\gamma'} } \lesssim 1 $  ), the cross-section can be estimated by using the Hulth$\acute{\text{e}}$n approximation~\cite{Tulin:2013teo}. 
Recently, the analytic formulas in the semi-classical regime ($ \frac{ \alpha'  m_{\chi} }{m_{\gamma'}} \gtrsim 1 $ and $ \frac{ m_{\chi} v_{\text{rel}} }{ m_{\gamma'} } \gtrsim 1 $) is also provided~\cite{Colquhoun:2020adl}, which fills the gap between the quantum regime and classical regime.

\begin{table}[htp]
\begin{center}
\begin{tabular}{ c c c }
\hline
\hline
System &\ \ \ scattering velocity $\langle v \rangle$  \ \ \ &  \ \ \  required $\bar{\sigma}/m_{\chi} $  \ \ \  \\ 
\hline
\hline
Dwarf galaxy / Galaxy  & \ \ \   10 - 200 km/s    \ \ \  & \ \ \  1 - 50 cm$^2$/g  \ \ \  \\
\hline
Galaxy groups & \ \ \  1150 km/s \ \ \  & \ \ \ 0.5$\pm$0.2 cm$^2$/g \ \ \  \\ 
\hline
Galaxy clusters & \ \ \  1900 km/s  \ \ \  & \ \ \   0.19$\pm$0.09 cm$^2$/g   \ \ \  \\ 
\hline
\end{tabular}
\caption{ Small scale data we considered to constrain DM scattering. }
\label{data}
\end{center}
\end{table}

\begin{figure}[ht]
\centering
\includegraphics[width=6.5in]{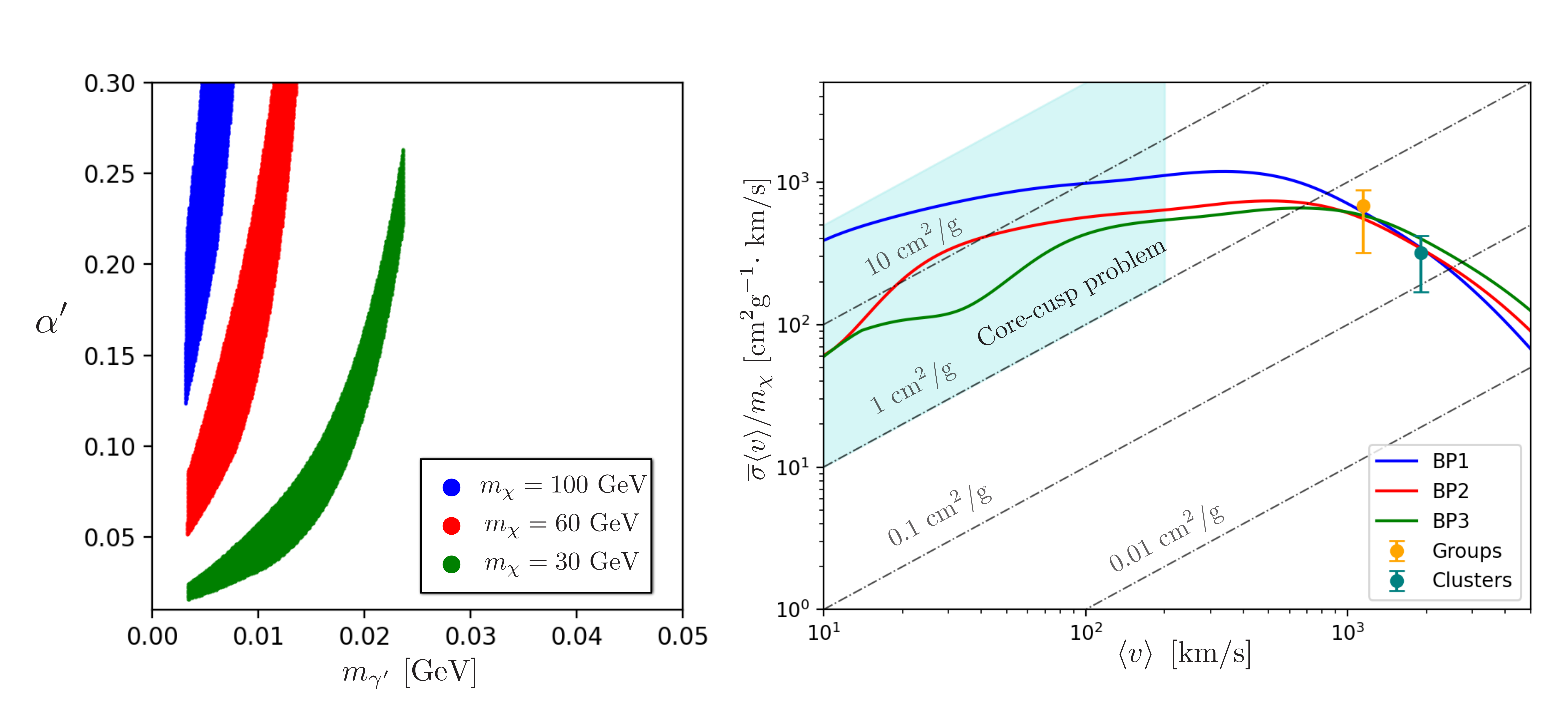}
\caption{  Left: parameter region consistent with small scale data with DM mass fixed to different values. 
Right: averaged elastic DM scattering cross-section as functions of DM scattering velocity for three benchmark points.  }
\label{SI}
\end{figure}

In the literature, momentum transfer cross section $\sigma_\text{T} \equiv \int d \Omega (1-\cos\theta) \frac{d\sigma}{d \Omega} $ is generally used as the proxy for DM elastic scattering. 
However, it is suggested to use viscosity cross section $\sigma_\text{V} \equiv \int d \Omega \sin^2\theta \frac{d\sigma}{d \Omega} $ instead of $\sigma_\text{T}$ as the proxy. Because $\sigma_\text{V}$ is more related to the heat conductivity and $\sigma_\text{V}$ is well defined for identical particles~\cite{Tulin:2013teo,GasD1965,Colquhoun:2020adl}.
Ref.~\cite{Colquhoun:2020adl} also suggest to use $ \overline{ \sigma_\text{V} } = \langle \sigma_\text{V} v^3_{\text{rel}}  \rangle / 24 \sqrt{\pi} v^3_0 $ as the velocity averaged cross section for this parameter is directly related to the energy transfer. 
All the above methods have been implemented in public code CLASSICS~\cite{Colquhoun:2020adl}, and we will use this code to calculate the DM elastic scattering cross-section in our model.

We list the observations we considered to constrain DM scattering in Tab.~\ref{data}. 
Fitting results for galaxy groups and clusters come from Ref.~\cite{Sagunski:2020spe}.
Firstly we perform a parameter scan with DM mass fixed to $100$ GeV, $60$ GeV, and $30$ GeV, respectively. 
The scan results are present in Fig.~\ref{SI} (left). 
It shows that for DM within mass range 10 GeV - 100 GeV, $\alpha' \sim  \mathcal{O}(0.1)$ and $m_{\gamma'} \sim $ $\mathcal{O}$(1) MeV -  $\mathcal{O}$(10) MeV 
are favored by small scale structure data. 
Coupling strength and mediator mass tend to decrease and increase respectively as DM become lighter.  
Furthermore, we choose three benchmark points to show the dependence of $\bar{\sigma}/m_{\chi}$ on scattering velocity: 
\begin{eqnarray}
\text{Benchmark Point 1: } &\ &  m_{\chi} = 100 \text{GeV} , \ m_{\gamma'} = 3.5 \text{MeV} , \ \alpha' =  0.15  \\\nonumber
\text{Benchmark Point 2: } &\ &  m_{\chi} = 60 \text{GeV} , \ m_{\gamma'} = 7 \text{MeV} , \ \alpha' = 0.1   \\\nonumber
\text{Benchmark Point 3: } &\ &  m_{\chi} = 30 \text{GeV} , \ m_{\gamma'} = 12 \text{MeV} , \ \alpha' =  0.05  
\end{eqnarray}
In Fig.~\ref{SI} (right) we present scattering cross section as functions of $\langle v \rangle$ for these three benchmark points.  
It shows a clear velocity dependence that fits the data. 

\section{Stochastic gravitational waves signal from dark $U(1)'$  phase transition \label{sec:PT} }

So for, we have built up a theory framework of ADM that is consistent with all the limits and can solve the small scale problems at the same time. 
In this section we discuss the detection of this scenario. 
Due to the nearly negligible portal between dark sector and visible sector in the scenario we chosen in this work, traditional methods are weak in detecting this scenario.  
But, if the spontaneous breaking of dark $U(1)'$ is induced by first order phase transition, then it is possible to detect the nearly independent dark sector by the stochastic gravitational wave signal~\cite{Witten:1984rs,Jaeckel:2016jlh,Schwaller:2015tja,Soni:2016yes,Addazi:2017gpt,Tsumura:2017knk,Huang:2017rzf,Hashino:2018zsi,Bai:2018dxf,Breitbach:2018ddu,Fairbairn:2019xog,Addazi:2020zcj,Ratzinger:2020koh,Ghosh:2020ipy,Dent:2022bcd,Wang:2022lxn,Wang:2022akn}. 
Here we perform a brief analysis.

The sector related to the MeV scale dark $U(1)'$ symmetry breaking is generally called Abelian Higgs model in the literature~\cite{Wainwright:2011qy,Chiang:2017zbz}. 
And Lattice simulation already shown that the phase transition of  Abelian Higgs model is first order, provided the Higgs mass is smaller or much smaller than gauge boson mass~\cite{Karjalainen:1996wx,Dimopoulos:1997cz}.
The corresponding Lagrangian is given by:
\begin{eqnarray}
\mathcal{L}_{U(1)'}  = -  \left(\partial_{\mu} S_2 + i 2 g' A'_{\mu} S_2  \right)^{\dagger} \left(\partial^{\mu} S_2 + i 2 g' A'^{\mu} S_2  \right) -\frac{1}{4} F'_{\mu\nu} F'^{\mu\nu} 
 +\mu_2^2 S_2^{\dagger} S_2 - \frac{\lambda_2}{4} \left( S_2^{\dagger} S_2 \right)^2
 \label{lagrangian}
\end{eqnarray}
Here we don not need to include $\chi$ and $S_1$ because their masses are far from MeV scale.  
The $U(1)'$ charge of $S_2$ has been fixed to +2 as we said in Sec.~\ref{sec:Model}.

After $S_2$ got VEV, it can be expressed as: 
\begin{eqnarray}
S_2 = \frac{1}{\sqrt{2}} ( v + s_2 + ia  )
\end{eqnarray}
Here $v \equiv 2 \mu_2 / \sqrt{\lambda_2} $ is the VEV of $S_2$ at zero temperature. $s_2$ and $a$ are the scalar and pseudo-scalar components of $S_2$ respectively. In $R_{\xi}$ gauge, the gauge-fixing and ghost terms are:
\begin{eqnarray}
\mathcal{L}_{\text{gf+gh}} = -\frac{1}{2} \xi^{-1} (\partial_{\mu} A^{\mu} - \xi 2g' v a )^2 - \bar{c} \left( -\partial_{\mu}\partial^{\mu} + \xi ( 2g' )^2 v(v+s_2)  \right) c
\end{eqnarray}
where $c$ is the ghost field. 
Zero temperature spectrum are given by: 
\begin{eqnarray}
 m^2_{s_2} =  \frac{1}{2} \lambda_2 v^2 \ , \ m^2_a = \xi (2g')^2 v^2 \ , \
 m^2_c = \xi (2g')^2 v^2 \ , \  m^2_{\gamma'} = (2g')^2 v^2 
 \label{mass1}
\end{eqnarray}

At finite temperature, $S_2$ field value $\phi \neq v$, and $\phi$ dependent spectrums are: 
\begin{eqnarray}
& & m^2_{s_2}(\phi) = \frac{3}{4}\lambda_2 \phi^2 - \mu^2_2      \ , \ 
 m^2_a(\phi) =  \frac{1}{4}\lambda_2 \phi^2 - \mu^2_2 + \xi (2g')^2 \phi^2 \ , \ \\\nonumber
& &  m^2_c(\phi) = \xi (2g')^2 \phi^2 \ , \  
 m^2_{\gamma'}(\phi) = (2g')^2 \phi^2 
 \label{mass2}
\end{eqnarray}
In the rest of this section we will consider Landau gauge ($\xi = 0$) to decouple ghost fields. 
$\{ m_{s_2},\ m_{\gamma'},\ \alpha' \}$ are chosen as input parameters to induce other relevant parameters.

\subsection{Thermal effective potential \label{sec:eft} }

Free energy density of dark $U(1)'$ sector is the thermal effective potential.
Thermal effective potential at temperature~\footnote{In this section, all the temperature labels represent dark sector temperature by default.} $T$ can be schematically expressed as: 
\begin{eqnarray}
V(\phi,T) = V^0(\phi) + V^{\text{1-loop}}(\phi) + V^\text{T}(\phi,T) + V^\text{daisy}(\phi,T)
\end{eqnarray}
Here $V^0$ is the tree-level potential, $V^{\text{1-loop}}$ is the sum of one-loop Coleman-Weinberg potential and counter-terms,
$V^{T}$ is the thermal correction, and $V^\text{daisy}$ is the correction from daisy resummation.

Tree-level potential comes from the potential sector of Lagrangian~(\ref{lagrangian}) by replacing $S_2$ by $\phi/\sqrt{2}$:
\begin{eqnarray}
V^0(\phi) = - \frac{1}{2}\mu^2_2 \phi^2 + \frac{1}{16} \lambda_2 \phi^4 
\end{eqnarray}

$V^{\text{1-loop}}$ is composed by one-loop Coleman-Weinberg potential and counter-terms, where the Coleman-Weinberg potential under $\overline{\text{MS}}$ renormalization scheme is~\cite{Coleman:1973jx} : 
\begin{eqnarray}
V^\text{CW}(\phi) &=& \frac{1}{64\pi^2} \Bigg\{  m_{s_2}^4(\phi) \left[ \ln\left( \dfrac{m_{s_2}^2(\phi)}{Q^2} \right)-\dfrac{3}{2}\right] 
    +  m_a^4(\phi) \left[ \ln\left( \dfrac{m_a^2(\phi)}{Q^2} \right)-\dfrac{3}{2}\right]   \\ \nonumber
  & & \ \ \ \ \ \  + 3 m_{\gamma'}^4(\phi) \left[ \ln\left( \dfrac{m_{\gamma'}^2(\phi)}{Q^2} \right)-\dfrac{5}{6}\right]     \Bigg\}
\end{eqnarray}
Here we need to emphasize that the potential parameter $\mu_2^2$ and $\lambda_2$ is determined by input physical parameters $\{m_{s_2}, m_{\gamma'}, \alpha' \}$ via tree-level relation Eq.~(\ref{mass1}).
Thus, to prevent physical mass and VEV being shifted by one-loop correction, 
counter terms need to be added to obey following on-shell conditions:
\begin{eqnarray}
\left. \frac{d }{ d \phi} \Delta V_{1}(\phi) \right|_{\phi=v} = 0 \ , \ \left. \frac{d^2 }{ d \phi^2} \Delta V_{1}(\phi) \right|_{\phi=v} = -\Delta \Sigma
  \label{onshell}
\end{eqnarray}
where $\Delta \Sigma \equiv \Sigma(m^2_{s_2}) - \Sigma(0) $ is the difference between scalar self-energy at different momentums. 
If all the involved particles are massive, it is harmless to ignore $\Delta \Sigma$ in Eq.~(\ref{onshell}).
But Goldstone $a$ in Landau gauge is massless, and it causes an infrared (IR) divergence when we perform on-shell conditions on Coleman-Weinberg potential. 
So we need the IR divergence in $\Delta \Sigma$ to make all IR divergences from Goldstone cancel out. 
See~\cite{Delaunay:2007wb} for more detailed discussion.
One-loop correction which satisfy (\ref{onshell}) is~\cite{Anderson:1991zb}: 
\begin{eqnarray}
V^{\text{1-loop}}(\phi) &=& \frac{1}{64\pi^2} \left[ m^4_{s_2}(\phi) \left( \ln \frac{m^2_{s_2}(\phi)}{m^2_{s_2}} -\frac{3}{2} \right) + 2 m^2_{s_2} m^2_{s_2}(\phi)  \right] \\\nonumber
 &+& \frac{3}{64\pi^2} \left[ m^4_{\gamma'}(\phi) \left( \ln \frac{m^2_{\gamma'}(\phi)}{m^2_{\gamma'}} -\frac{3}{2} \right) + 2 m^2_{\gamma'} m^2_{\gamma'}(\phi)  \right] \\\nonumber
 &+& \frac{1}{64\pi^2} \left[ m^4_a(\phi) \left( \ln \frac{m^2_a(\phi)}{m^2_{s_2}} -\frac{3}{2} \right)  \right]
  \label{CW}
\end{eqnarray}

Thermal correction is~\cite{thermal_correction1,thermal_correction2}:
\begin{eqnarray}
V^\text{T}(\phi,T) = \frac{T'^4}{2\pi^2} \Bigg[  J_B\left( \dfrac{m_{s_2}^2(\phi)}{T^2} \right)  
    + J_B\left( \dfrac{m_a^2(\phi)}{T^2} \right) 
    + 3 J_B\left( \dfrac{m_{\gamma'}^2(\phi)}{T^2} \right)  \Bigg]
\end{eqnarray}
Here the bosonic thermal function $J_B$ is:
\begin{eqnarray}
J_B(x) = \int^{\infty}_0 k^2 \ln\left[1- \exp\left( -\sqrt{k^2 + x} \right)  \right]  dk
\end{eqnarray}

To avoid the IR divergence when the mass of boson is much smaller than temperature, daisy resummation needs to be added for scalar and longitudinal component of $\gamma'$~\cite{Arnold:1992rz} :
\begin{eqnarray}
V^\text{daisy}(\phi,T) &=& - \frac{T}{12\pi} \Bigg\{  \left[ \left(  {m}^2_{s_2}(\phi) + \Pi_{s_2}(T) \right)^{\frac{3}{2}} - \left(  {m}^2_{s_2}(\phi)  \right)^{\frac{3}{2}} \right]  \\\nonumber
 & & \ \ \ \ \  + \left[ \left( {m}^2_a(\phi)   + \Pi_{a}(T)  \right)^{\frac{3}{2}} - \left(  {m}^2_a(\phi) \right)^{\frac{3}{2}} \right]  + \left[ \left( {m}^2_{\gamma'}(\phi)  + \Pi_{\gamma'}(T)  \right)^{\frac{3}{2}} - \left( {m}^2_{\gamma'}(\phi)  \right)^{\frac{3}{2}} \right]      \Bigg\}
\end{eqnarray}
where $\Pi_{s_2}(T) = \left( \lambda_2/6 + (2g')^2/4 \right)T^2 $, 
$\Pi_{a}(T) = \left( \lambda_2/6 + (2g')^2/4 \right)T^2$, and 
$\Pi_{\gamma'}(T) = \left(  (2g')^2/3 \right)T^2$ are thermal Debye mass squares.

\subsection{Nucleation temperature \label{sec:nc} }

When the temperature is below the critical temperature, false vacuum transfer to true vacuum via thermally fluctuation~\footnote{In the case of dark sector being much colder than visible sector, quantum tunneling also need to be considered. See Ref.~\cite{Fairbairn:2019xog} for detailed discussion.}. 
Transition rate per unit volume  is given by~\cite{Coleman:1977py,Callan:1977pt,Linde:1981zj}: 
\begin{eqnarray}
\Gamma(T) = A(T) e^{-S_E(T)}
\end{eqnarray}
Here $A(T)=\omega T^4$ with $\omega\sim \mathcal{O}(1)$, and $S_E(T)$ is the 4-D Euclidean action. 
In the case of thermal transition, $S_E(T)$ is the ratio between 3-D Euclidean action $S_3(T)$ and temperature:  
\begin{eqnarray}
S_E(T) = \frac{S_3(T)}{T}
\end{eqnarray}
And 3-D Euclidean action $S_3(T)$ is given by: 
\begin{eqnarray}
S_3(T) = \int d^3 x \left[ \frac{1}{2} (\nabla \phi)^2 + V(\phi,T) \right] 
\end{eqnarray}
Due to the 3-D rotation invariance, $\phi$ only depend on radius $r$ and thus $S_3$ can be rewritten as:   
\begin{eqnarray}
S_3(T) = 4\pi \int_0^{\infty} r^2 dr \left[ \frac{1}{2} \left( \frac{d\phi}{dr} \right)^2 + V(\phi,T) \right] 
\end{eqnarray}
Minimization condition of $S_3(T)$ gives the equation of motion that $\phi$ should follow: 
\begin{eqnarray}
\frac{d^2\phi}{d r^2} + \frac{2}{r} \frac{d\phi}{d r} = \frac{\partial V(\phi,T)}{\partial \phi}
\label{Eom}
\end{eqnarray}
Adding boundary conditions $\lim\limits_{r \to \infty} \phi(r) = 0$ and $\left. \frac{d\phi}{dr} \right|_{r=0} = 0$, Eq.~(\ref{Eom}) can be solved numerically by overshoot/undershoot method~\cite{Apreda:2001us}. In this work we use public code CosmoTransitions~\cite{Wainwright:2011kj} to do the calculation. 

Nucleation starts at the temperature where the transition rate within one Hubble volume approximates Hubble rate: 
\begin{eqnarray}
& & H(T_N)^{-3} \times \Gamma(T_N) \approx H(T_N) \\\nonumber 
\Rightarrow & &  A(T_N) e^{-  {S_3(T_N)}/{T_N}  } \approx \left(  1.66\frac{ T_N^2 }{M_{\text{Pl}}} \right)^4 ( g_{\star}/\xi^4  +  g'_{\star} )^2 \\\nonumber 
\Rightarrow & & \frac{S_3(T_N)}{T_N}  \approx   4\ln \left(  \frac{ M_{\text{Pl}} }{T_N}  \right) - 2 \ln ( g_{\star}/\xi^4  +  g'_{\star} ) - 2.027
\end{eqnarray}
Here $T_N$ is the nucleation temperature, and we approximate $\omega$ to 1 in the third line.
In our model, phase transition in the dark $U(1)'$ sector happens around MeV scale, and the temperature ratio $\xi$ is generally not much smaller than 1. 
So the nucleation temperature $T_N$ is approximately determined by ${S_3(T_N)}/{T_N} \approx 196 $.

\begin{table}[htp]
\begin{center}
\begin{tabular}{ c c c c  c c c }
\hline
\hline
Benchmark point &\ $m_{\chi}$ (GeV) \ &\ $m_{\gamma'}$ (MeV) \ &\ $m_{s_2}$ (MeV) \ &\ $\alpha'$ \ &\ $\xi_{\text{ini}}$ \ &\  $T_N$ (MeV) \   \\ 
\hline
\hline
BP1  & \ 100 \  & \ 3.5 \  & \ 1.5 \  & \ 0.15 \  & \ 0.7 \  &  \ 1.63 \   \\
\hline
BP2  & \ 60 \  & \ 7 \  & \ 2.5 \  & \ 0.1 \  & \ 0.7 \  &  \ 1.83 \   \\
\hline
BP3  & \ 30 \  & \ 12 \  & \ 3.8 \  & \ 0.05 \  & \ 0.7 \  &  \ 3.75 \   \\
\hline
\end{tabular}
\caption{Benchmark points we considered for first order phase transition study. }
\label{bp}
\end{center}
\end{table}

In Tab.~\ref{bp} we present three benchmark points for illustration in this section. 
These three benchmark points are also consistent with small scale structure data.

\subsection{Phase Transition parameters \label{sec:pt} }

After nucleation, bubbles expand rapidly and after a while collide with each other and generate gravitational waves (GWs). 
There are three GWs generation mechanisms: bubble walls collision~\cite{Kosowsky:1991ua,Kosowsky:1992rz,Kosowsky:1992vn,Kamionkowski:1993fg,Caprini:2007xq,Huber:2008hg}, sound waves~\cite{Hindmarsh:2013xza,Giblin:2013kea,Giblin:2014qia,Hindmarsh:2015qta}, and magnetohydrodynamic turbulence~\cite{Caprini:2006jb,Kahniashvili:2008pf,Kahniashvili:2008pe,Kahniashvili:2009mf,Caprini:2009yp,Kisslinger:2015hua}. 
The generated gravitational waves stay in the universe and redshift in wavelength as the universe expands. 
To obtain current spectrum of these phase transition gravitational waves, firstly we need to calculate a set of parameters used to describe the phase transition dynamics: 
$T_\ast$, $\mathbb{A}$~\footnote{To avoid confusion with the label of fine structure constant, in this paper we use $\mathbb{A}$ to represent the strength parameter.}, $\mathbb{A}'$ ($\mathbb{A}$ in dark sector), $\beta/H_\ast$, $v_w$, and $\kappa_{b,s,t}$. 
Meaning of these parameters are given below. 

$T_\ast$ is the characteristic temperature of GWs generation. 
Generally, $T_\ast$ can be chosen as percolation temperature, the temperature at which a large fraction of the space has been occupied by bubbles~\cite{Ellis:2018mja,Ellis:2020awk,Wang:2020jrd}. 
But in the weak or mild supercooling case, using nucleation temperature $T_N$ as $T_\ast$ is also a good approximation. 
For the benchmark points we will study in this section, it is fine to approximates $T_\ast$ by $T_N$ because of their mild super cooling. 

Strength parameter $\mathbb{A}$ is the change in the trace of energy-momentum tensor during phase transition divided by relativistic energy density:
\begin{eqnarray}
\mathbb{A} = \frac{1}{\rho_\ast} \left[ \Delta V - \frac{T}{4}\frac{d \Delta V}{d T} \right]_{T=T_\ast}
\end{eqnarray}
Here $\rho_\ast$ is the relativistic energy density at $T_\ast$. $\Delta V$ is the difference in free energy density between false vacuum and true vacuum.  

It will be convenient to study dark sector dynamics if we define another strength parameter $\mathbb{A}'$ by only considering the relativistic energy density in the dark sector~\cite{Fairbairn:2019xog}: 
\begin{eqnarray}
\mathbb{A}' = \frac{1}{\rho'_\ast} \left[ \Delta V - \frac{T}{4} \frac{d \Delta V}{d T} \right]_{T=T_\ast} = \frac{\rho_\ast}{\rho'_\ast} \mathbb{A}
\end{eqnarray}
Here $\rho'_\ast$ the relativistic energy density in the dark sector at $T_\ast$. 

$\beta$ is the inverse of the duration of phase transition. Its ratio to Hubble expansion rate at $T_\ast$ is given by: 
\begin{eqnarray}
\frac{\beta}{H_{\ast}} = T_\ast \left. \frac{d S_E }{d T} \right|_{T=T_\ast}
\end{eqnarray}

$v_w$ is the velocity of bubble wall. 
$\kappa_{b,s,t}$ are the fractions of released vacuum energy that transferred to scalar-field gradients, sound waves, and turbulence, respectively.
Before estimating these parameters, we need to judge whether the phase transition is ``runaway'' or ``non-runaway''.
To do that, firstly we calculate the so-called threshold value of $\mathbb{A}'$, which is labeled as $\mathbb{A}'_{\infty}$~\cite{Espinosa:2010hh}: 
\begin{eqnarray}
\mathbb{A}'_{\infty} = \frac{1}{\rho'_\ast } \frac{T^2_\ast}{24} \left( \sum_i c_i n_i \Delta m_i^2  \right)
\end{eqnarray}
Here $c_i = 1 \ (1/2)$ for bosons (fermions), $n_i$ is the number of degrees of freedom (absolute value), and $\Delta m_i^2$ is the difference in particle mass square between false vacuum and true vacuum.

If $\mathbb{A}' > \mathbb{A}'_{\infty}$, the driving pressure will be larger than the friction from dark plasma and thus the bubble wall will eventually be accelerated to the maximal value, i.e. $v_w = 1$. 
This is the so-called "runaway" case. 
In this case, fractions $\kappa_{b,s,t}$ are given by~\cite{Hindmarsh:2015qta,Schmitz:2020syl}: 
\begin{eqnarray}
\kappa_b = 1 - \frac{\mathbb{A}'_{\infty}}{\mathbb{A}'} \ , \ \kappa_s = \frac{\mathbb{A}'_{\infty}}{\mathbb{A}'} \frac{\mathbb{A}'_{\infty}}{0.73+0.083\sqrt{\mathbb{A}'_{\infty}} + \mathbb{A}'_{\infty} }  \ , \ \kappa_t = 0.1\kappa_s
\end{eqnarray}

If $\mathbb{A}' < \mathbb{A}'_{\infty}$, bubble wall will eventually reach a subluminal velocity and this is called "non-runaway" phase transition. 
In this case we simply choice $v_w=0.9$ for a fast and rough estimation of GWs signal~\footnote{The calculation of $v_w$ in a concrete model is still quite difficult. See Ref.~\cite{Moore:1995si,Megevand:2009gh,Huber:2013kj,Konstandin:2014zta,Dorsch:2018pat,Laurent:2022jrs,Wang:2020zlf} for previous studies.}. 
For non-runaway phase transition, the main source of GWs will be sound waves and the contribution from bubble collision is negligible. 
Fractions $\kappa_{b,s,t}$ are given by~\cite{Espinosa:2010hh}: 
\begin{eqnarray}
\kappa_b \simeq 0 \ , \ \kappa_s =  \frac{\mathbb{A}'}{0.73+0.083\sqrt{\mathbb{A}'} + \mathbb{A}' }  \ , \ \kappa_t = 0.1\kappa_s
\end{eqnarray}

\begin{table}[htp]
\begin{center}
\begin{tabular}{ c c c c c c c c c c}
\hline
\hline
Benchmark point &\ $T_N$ (MeV)  \ &\ $\mathbb{A}$  \ &\ $\mathbb{A}'$  \ &\ $\mathbb{A}'_{\infty}$ \ &\ ${\beta}/{H_{\ast}}$ \ &\ $v_w$ \ &\ $\kappa_b$  \ &\  $\kappa_s$  \ &\  $\kappa_t$ \    \\ 
\hline
\hline
BP1  & \ 1.63 \  & \ $3.20\times 10^{-4}$ \  & \ 0.0234 \  & \ 0.410 \  & \ 3378.9 \  & \ 0.9 \  & \ 0 \ & \ 0.0305 \ & \ 0.00305 \    \\
\hline
BP2  & \ 1.83 \  & \ $1.61\times 10^{-3}$ \  & \ 0.117 \  & \ 1.05 \  & \ 784.7 \  & \ 0.9 \  & \ 0 \ & \ 0.134 \ & \ 0.0134  \     \\
\hline
BP3  & \ 3.75 \  & \ $1.79\times 10^{-3}$ \  & \ 0.131 \  & \ 0.657 \  & \ 2060.8 \  & \ 0.9 \  & \ 0 \ & \ 0.147 \ & \ 0.0147 \     \\
\hline
\end{tabular}
\caption{ Phase transition parameters.  }
\label{pt}
\end{center}
\end{table}

In Tab.~\ref{pt} we present all the phase transition parameters for the three benchmark points. 

\subsection{Gravitational waves \label{sec:gw} }

In this subsection we present the calculation of today's GWs signal in our model. 
formulas used in this subsection can be found in the literature~\cite{Caprini:2015zlo,Caprini:2019egz,Huber:2008hg,Hindmarsh:2015qta,Caprini:2009yp}. 

The total GWs signal is the linear superposition of spectrums from bubble collisions, sound waves, and turbulence:
\begin{eqnarray}
h^2\Omega_{\text{GW}} (f)  = h^2\Omega_{{b}} (f)  + h^2\Omega_{{s}} (f) + h^2 \Omega_{{t}} (f) 
\end{eqnarray}
The three indivisual contributions can be further divided into peak amplitudes ($\Omega^{\text{peak}}_{{b,s,t}}$) and  spectral shape functions ($\mathcal{S}_{{b,s,t}}$): 
\begin{eqnarray}
& & h^2\Omega_{{b}} (f) \simeq h^2 \Omega^{\text{peak}}_{{b}} \mathcal{S}_{{b}}(f,f^{\text{peak}}_{{b}}) \\\nonumber
& & h^2\Omega_{{s}} (f) \simeq h^2 \Omega^{\text{peak}}_{{s}} \mathcal{S}_{{s}}(f,f^{\text{peak}}_{{s}})  \\\nonumber
& & h^2\Omega_{{t}} (f) \simeq h^2 \Omega^{\text{peak}}_{{t}} \mathcal{S}_{{t}}(f,f^{\text{peak}}_{{t}})
\end{eqnarray}
Peak amplitudes are determined by all the phase transition parameters we obtained before: 
\begin{eqnarray}
& & h^2 \Omega^{\text{peak}}_{{b}} = 1.24\times 10^{-5} \left( \frac{  {h_{\text{eff}}}_0   }{ {h_{\text{eff}}}_\ast  } \right)^{4/3} ({g_{\text{eff}}}_\ast) (\xi_0)^4  \left( \frac{0.11 v_w}{0.42+v_w^2} \right) \left( \frac{\kappa_b \mathbb{A}}{ 1+\mathbb{A} } \right)^2 \left( \frac{v_w}{\beta/H_\ast} \right)^2  \\\nonumber
& & h^2 \Omega^{\text{peak}}_{{s}} = 1.97\times 10^{-6} \left( \frac{  {h_{\text{eff}}}_0   }{ {h_{\text{eff}}}_\ast  } \right)^{4/3} ({g_{\text{eff}}}_\ast) (\xi_0)^4    \left( \frac{\kappa_s \mathbb{A}}{ 1+\mathbb{A} } \right)^2 \left( \frac{v_w}{\beta/H_\ast} \right)    \\\nonumber
& & h^2 \Omega^{\text{peak}}_{{t}} = 2.49\times 10^{-4} \left( \frac{  {h_{\text{eff}}}_0   }{ {h_{\text{eff}}}_\ast  } \right)^{4/3} ({g_{\text{eff}}}_\ast) (\xi_0)^4    \left( \frac{\kappa_t \mathbb{A}}{ 1+\mathbb{A} } \right)^{3/2} \left( \frac{v_w}{\beta/H_\ast} \right)
\end{eqnarray}
where ``$\ast$'' and ``$0$'' correspond to GWs producing time and current time, respectively. 
These formulas look different from expressions commonly found in the literature, because we need to recalculate the redshift factor for MeV scale dark phase transition~\cite{Breitbach:2018ddu}. 
For our benchmark points, the factor inside above expressions can be approximated to: 
\begin{eqnarray}
\left( \frac{  {h_{\text{eff}}}_0   }{ {h_{\text{eff}}}_\ast  } \right)^{4/3} ({g_{\text{eff}}}_\ast) (\xi_0)^4 \simeq \left( \frac{  58.7   }{ 173.2  } \right)^{4/3} \times 424.0 \times (0.41)^4
= 2.83
\end{eqnarray}

Spectral shape functions are given by: 
\begin{eqnarray}
& &  \mathcal{S}_b(f, f^{\text{peak}}_b )  =  \left( \frac{f}{f^{\text{peak}}_{{b}}} \right)^{2.8} \left[ \frac{3.8}{1 + 2.8 (f/f^{\text{peak}}_{{b}})^{3.8}} \right]  \\\nonumber
& &  \mathcal{S}_s(f, f^{\text{peak}}_s )  =  \left( \frac{f}{f^{\text{peak}}_{{s}}} \right)^{3} \left[ \frac{7}{4 + 3 (f/f^{\text{peak}}_{{s}})^{2}} \right]^{7/2}   \\\nonumber
& &  \mathcal{S}_t(f, f^{\text{peak}}_t )  = \left( \frac{f}{f^{\text{peak}}_{{t}}} \right)^{3} \left[ \frac{1}{1 +  (f/f^{\text{peak}}_{{t}})} \right]^{11/3} \frac{1}{1 + 8\pi f/h_\ast }
\end{eqnarray}
where: 
\begin{eqnarray}
h_\ast = \frac{a_\ast}{a_0} H_\ast = 7.44\times 10^{-11} \text{Hz} \left( \frac{   {g_{\text{eff}}}^{1/2}_\ast  }{ {h_{\text{eff}}}^{1/3}_\ast  } \right)  \left( \frac{  {h_{\text{eff}}}_0  }{ 3.91  } \right)^{1/3} \xi_0 \left( \frac{ T_\ast }{1 \text{ MeV} } \right) 
\end{eqnarray}
where 3.91 is current degree of freedom for entropy in the visible sector.  
For the MeV scale dark phase transition in our model, the value of $h_\ast$ can be approximated to: 
\begin{eqnarray}
h_\ast \simeq 2.78 \times 10^{-10} \text{Hz} \left( \frac{ T_\ast }{1 \text{ MeV} } \right) 
\end{eqnarray}
And peak frequencies are given by:
\begin{eqnarray}
 f^{\text{peak}}_b &=& 7.44\times 10^{-11} \text{Hz} \left( \frac{   {g_{\text{eff}}}^{1/2}_\ast  }{ {h_{\text{eff}}}^{1/3}_\ast  } \right)  \left( \frac{  {h_{\text{eff}}}_0  }{ 3.91  } \right)^{1/3} \xi_0 \left( \frac{ T_\ast }{1 \text{ MeV} } \right) \left( \frac{ \beta/H_\ast }{v_w} \right) \left( \frac{0.62 v_w}{ 1.8 - 0.1 v_w + v_w^2 } \right)  \\\nonumber
&\simeq& 2.78 \times 10^{-10} \text{Hz} \left( \frac{ T_\ast }{1 \text{ MeV} } \right) \left( \frac{ \beta/H_\ast }{v_w} \right) \left( \frac{0.62 v_w}{ 1.8 - 0.1 v_w + v_w^2 } \right)   \end{eqnarray}
\begin{eqnarray}
f^{\text{peak}}_s &=& 8.59\times 10^{-11} \text{Hz} \left( \frac{   {g_{\text{eff}}}^{1/2}_\ast  }{ {h_{\text{eff}}}^{1/3}_\ast  } \right)  \left( \frac{  {h_{\text{eff}}}_0  }{ 3.91  } \right)^{1/3} \xi_0 \left( \frac{ T_\ast }{1 \text{ MeV} } \right) \left( \frac{ \beta/H_\ast }{v_w} \right)   \\\nonumber
  &\simeq& 3.21 \times 10^{-10} \text{Hz} \left( \frac{ T_\ast }{1 \text{ MeV} } \right) \left( \frac{ \beta/H_\ast }{v_w} \right)  
\end{eqnarray}
\begin{eqnarray}
f^{\text{peak}}_t &=& 13.0\times 10^{-11} \text{Hz} \left( \frac{   {g_{\text{eff}}}^{1/2}_\ast  }{ {h_{\text{eff}}}^{1/3}_\ast  } \right)  \left( \frac{  {h_{\text{eff}}}_0  }{ 3.91  } \right)^{1/3} \xi_0 \left( \frac{ T_\ast }{1 \text{ MeV} } \right) \left( \frac{ \beta/H_\ast }{v_w} \right) \\\nonumber
 &\simeq& 4.87 \times 10^{-10} \text{Hz} \left( \frac{ T_\ast }{1 \text{ MeV} } \right) \left( \frac{ \beta/H_\ast }{v_w} \right)
\end{eqnarray}

\begin{figure}[ht]
\centering
\includegraphics[width=6.5in]{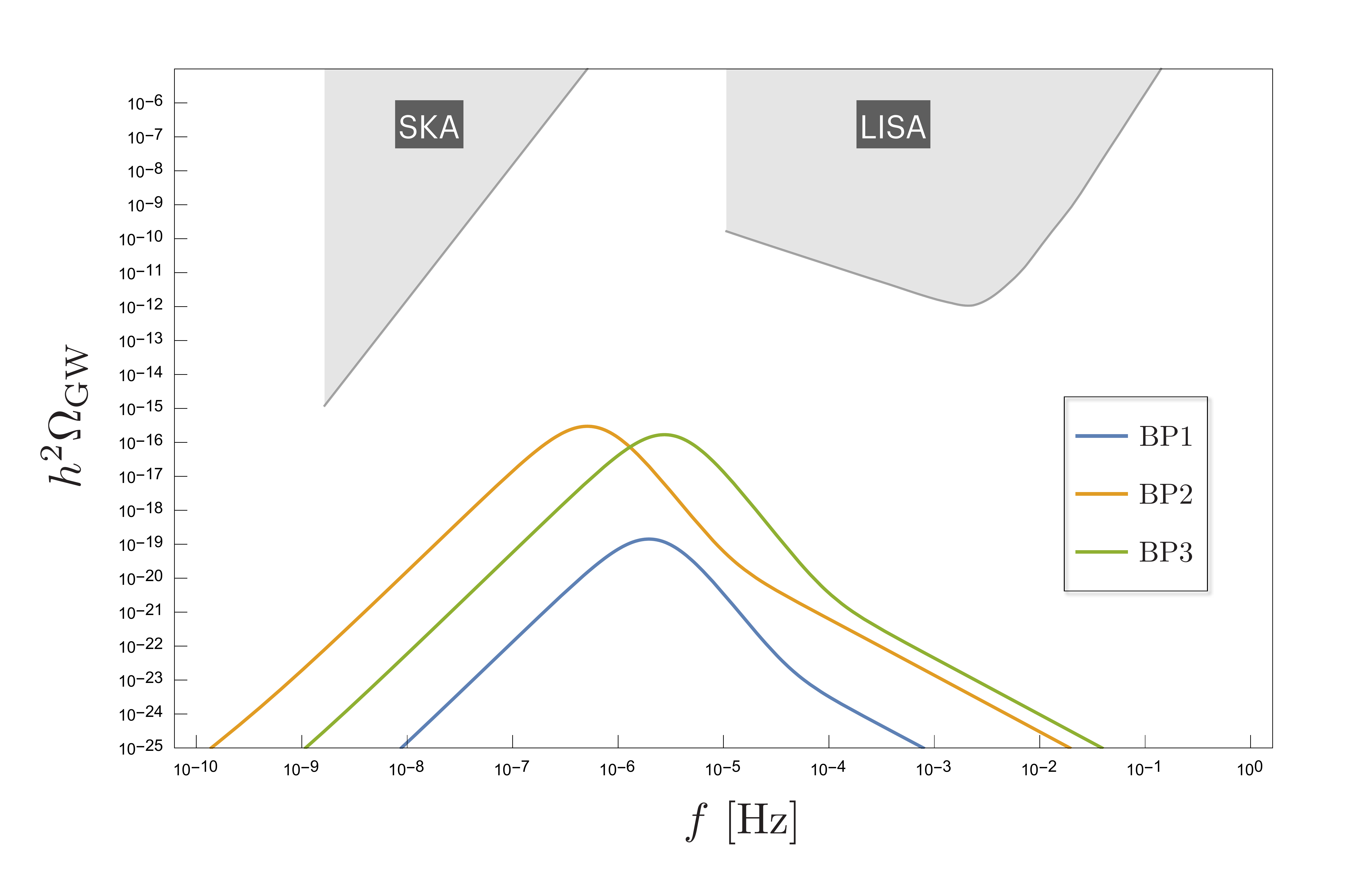}
\caption{ MeV scale $U(1)'$ phase transition GWs spectrums for three benchmark points. Gray areas are detection regions of SKA and LISA. }
\label{GW}
\end{figure}

For our three benchmark points, factor $\left( \frac{ \beta/H_\ast }{v_w} \right) \sim \mathcal{O}(1000)$.
Thus their peak frequencies are around $10^{-6} - 10^{-7}$ Hz, which is not favored by either SKA telescope~\cite{Janssen:2014dka} or space-based LISA interferometer~\cite{LISA:2017pwj}. 
In Fig.~\ref{GW} we present the GWs spectrums of our three benchmark points. 
As we expected, these signals are barely detectable by the SAK or LISA. 

However, this result depends significantly on our choice of benchmark point. 
In order not to change the limit about $N_{\text{eff}}$ we got in Sec.~\ref{sec:neff}, 
for all the benchmark points we consider, the strength of phase transitions are quite weak (i.e. the value of $\mathbb{A}$ and $\mathbb{A}'$ are quite small). 
If we increase the intensity of phase transition and make a strong supercooling, then the generated GWs signal can be easily detected by SKA. 
The reason is twofold. 
Firstly, $h^2 \Omega_{\text{GW}}$ is approximately proportional to $\mathbb{A}^2$. 
So increasing $\mathbb{A}$ by an order of magnitude, we can increase $h^2 \Omega_{\text{GW}}$ by roughly two orders of magnitude. 
Secondly, strong supercooling makes $T_{\ast}$ far below MeV scale, and thus make the peak frequency of $h^2 \Omega_{\text{GW}}$ closer to the detection region of SKA.
Certainly, in the strong supercooling case we need to revisit the limit from $N_{\text{eff}}$. 
Detailed analysis is left for a future study.

\section{Conclusion}\label{sec:conclusion}

In this work we propose an asymmetry DM model with massive mediator to explain DM small scale structure data and to avoid the limit from CMB. 
In our model, the DM candidate is a vector-like fermion charged under a dark $U(1)'$, and the mediator is the $U(1)'$ gauge boson that gain mass from the spontaneous symmetry breaking.  
The asymmetry between DM and anti-DM is generated by the CP violated and out-of-equilibrium decay of a neutral heavy fermion. 
The model is consistent with cosmology observations like CMB and LLS. 
The existence of dark radiation increases the value of $N_{\text{eff}}$, and it makes this model to be detectable by the future measurement of $N_{\text{eff}}$.
Finally, the MeV scale $U(1)'$ symmetry breaking generate GWs signal with peak frequency around $10^{-6} - 10^{-7} \text{ Hz}$.  
It also possible to make the GWs from $U(1)'$ symmetry breaking to be detected by SKA, if we consider a strong supercooling phase transition.

\section*{Acknowledgements}
M.Z. thanks Song Li and Yang Xiao for useful discussions. 
This work was supported by the National Natural Science Foundation of China (NNSFC) under grant No. 12105118 and 11947118.

\vspace{0.5cm}

\bibliographystyle{apsrev4-1}

\begin{thebibliography}{999}


\bibitem{Ade:2015xua} 
  P.~A.~R.~Ade {\it et al.} [Planck Collaboration],
``Planck 2015 results. XIII. Cosmological parameters,''
  Astron.\ Astrophys.\  {\bf 594}, A13 (2016)
  doi:10.1051/0004-6361/201525830
  [arXiv:1502.01589 [astro-ph.CO]].

\bibitem{Clowe:2006eq}
D.~Clowe, M.~Bradac, A.~H.~Gonzalez, M.~Markevitch, S.~W.~Randall, C.~Jones and D.~Zaritsky,
``A direct empirical proof of the existence of dark matter,''
Astrophys. J. Lett. \textbf{648}, L109-L113 (2006)
doi:10.1086/508162
[arXiv:astro-ph/0608407 [astro-ph]].


\bibitem{Blumenthal:1984bp}
G.~R.~Blumenthal, S.~M.~Faber, J.~R.~Primack and M.~J.~Rees,
``Formation of Galaxies and Large Scale Structure with Cold Dark Matter,''
Nature \textbf{311}, 517-525 (1984)
doi:10.1038/311517a0

\bibitem{Springel:2006vs}
V.~Springel, C.~S.~Frenk and S.~D.~M.~White,
``The large-scale structure of the Universe,''
Nature \textbf{440}, 1137 (2006)
doi:10.1038/nature04805
[arXiv:astro-ph/0604561 [astro-ph]].

\bibitem{Flores:1994gz}
R.~A.~Flores and J.~R.~Primack,
``Observational and theoretical constraints on singular dark matter halos,''
Astrophys. J. Lett. \textbf{427}, L1-4 (1994)
doi:10.1086/187350
[arXiv:astro-ph/9402004 [astro-ph]].

\bibitem{Moore:1994yx}
B.~Moore,
``Evidence against dissipationless dark matter from observations of galaxy haloes,''
Nature \textbf{370}, 629 (1994)
doi:10.1038/370629a0

\bibitem{Moore:1999gc}
B.~Moore, T.~R.~Quinn, F.~Governato, J.~Stadel and G.~Lake,
``Cold collapse and the core catastrophe,''
Mon. Not. Roy. Astron. Soc. \textbf{310}, 1147-1152 (1999)
doi:10.1046/j.1365-8711.1999.03039.x
[arXiv:astro-ph/9903164 [astro-ph]].


\bibitem{Oman:2015xda}
K.~A.~Oman, J.~F.~Navarro, A.~Fattahi, C.~S.~Frenk, T.~Sawala, S.~D.~M.~White, R.~Bower, R.~A.~Crain, M.~Furlong and M.~Schaller, \textit{et al.}
``The unexpected diversity of dwarf galaxy rotation curves,''
Mon. Not. Roy. Astron. Soc. \textbf{452}, no.4, 3650-3665 (2015)
doi:10.1093/mnras/stv1504
[arXiv:1504.01437 [astro-ph.GA]].


\bibitem{Klypin:1999uc}
A.~A.~Klypin, A.~V.~Kravtsov, O.~Valenzuela and F.~Prada,
``Where are the missing Galactic satellites?,''
Astrophys. J. \textbf{522}, 82-92 (1999)
doi:10.1086/307643
[arXiv:astro-ph/9901240 [astro-ph]].

\bibitem{Moore:1999nt}
B.~Moore, S.~Ghigna, F.~Governato, G.~Lake, T.~R.~Quinn, J.~Stadel and P.~Tozzi,
``Dark matter substructure within galactic halos,''
Astrophys. J. Lett. \textbf{524}, L19-L22 (1999)
doi:10.1086/312287
[arXiv:astro-ph/9907411 [astro-ph]].


\bibitem{Boylan-Kolchin:2011qkt}
M.~Boylan-Kolchin, J.~S.~Bullock and M.~Kaplinghat,
``Too big to fail? The puzzling darkness of massive Milky Way subhaloes,''
Mon. Not. Roy. Astron. Soc. \textbf{415}, L40 (2011)
doi:10.1111/j.1745-3933.2011.01074.x
[arXiv:1103.0007 [astro-ph.CO]].

\bibitem{Boylan-Kolchin:2011lmk}
M.~Boylan-Kolchin, J.~S.~Bullock and M.~Kaplinghat,
``The Milky Way's bright satellites as an apparent failure of LCDM,''
Mon. Not. Roy. Astron. Soc. \textbf{422}, 1203-1218 (2012)
doi:10.1111/j.1365-2966.2012.20695.x
[arXiv:1111.2048 [astro-ph.CO]].

\bibitem{Navarro:1996bv}
J.~F.~Navarro, V.~R.~Eke and C.~S.~Frenk,
``The cores of dwarf galaxy halos,''
Mon. Not. Roy. Astron. Soc. \textbf{283}, L72-L78 (1996)
doi:10.1093/mnras/283.3.L72
[arXiv:astro-ph/9610187 [astro-ph]].

\bibitem{Governato:2009bg}
F.~Governato, C.~Brook, L.~Mayer, A.~Brooks, G.~Rhee, J.~Wadsley, P.~Jonsson, B.~Willman, G.~Stinson and T.~Quinn, \textit{et al.}
``At the heart of the matter: the origin of bulgeless dwarf galaxies and Dark Matter cores,''
Nature \textbf{463}, 203-206 (2010)
doi:10.1038/nature08640
[arXiv:0911.2237 [astro-ph.CO]].



\bibitem{Spergel:1999mh}
D.~N.~Spergel and P.~J.~Steinhardt,
``Observational evidence for selfinteracting cold dark matter,''
Phys. Rev. Lett. \textbf{84}, 3760-3763 (2000)
doi:10.1103/PhysRevLett.84.3760
[arXiv:astro-ph/9909386 [astro-ph]].



\bibitem{Rocha:2012jg}
M.~Rocha, A.~H.~G.~Peter, J.~S.~Bullock, M.~Kaplinghat, S.~Garrison-Kimmel, J.~Onorbe and L.~A.~Moustakas,
``Cosmological Simulations with Self-Interacting Dark Matter I: Constant Density Cores and Substructure,''
Mon. Not. Roy. Astron. Soc. \textbf{430}, 81-104 (2013)
doi:10.1093/mnras/sts514
[arXiv:1208.3025 [astro-ph.CO]].

\bibitem{Peter:2012jh}
A.~H.~G.~Peter, M.~Rocha, J.~S.~Bullock and M.~Kaplinghat,
``Cosmological Simulations with Self-Interacting Dark Matter II: Halo Shapes vs. Observations,''
Mon. Not. Roy. Astron. Soc. \textbf{430}, 105 (2013)
doi:10.1093/mnras/sts535
[arXiv:1208.3026 [astro-ph.CO]].

\bibitem{Moore:2000fp}
B.~Moore, S.~Gelato, A.~Jenkins, F.~R.~Pearce and V.~Quilis,
``Collisional versus collisionless dark matter,''
Astrophys. J. Lett. \textbf{535}, L21-L24 (2000)
doi:10.1086/312692
[arXiv:astro-ph/0002308 [astro-ph]].

\bibitem{Yoshida:2000bx}
N.~Yoshida, V.~Springel, S.~D.~M.~White and G.~Tormen,
``Collisional dark matter and the structure of dark halos,''
Astrophys. J. Lett. \textbf{535}, L103 (2000)
doi:10.1086/312707
[arXiv:astro-ph/0002362 [astro-ph]].

\bibitem{Burkert:2000di}
A.~Burkert,
``The Structure and evolution of weakly selfinteracting cold dark matter halos,''
Astrophys. J. Lett. \textbf{534}, L143-L146 (2000)
doi:10.1086/312674
[arXiv:astro-ph/0002409 [astro-ph]].

\bibitem{Kochanek:2000pi}
C.~S.~Kochanek and M.~J.~White,
``A Quantitative study of interacting dark matter in halos,''
Astrophys. J. \textbf{543}, 514 (2000)
doi:10.1086/317149
[arXiv:astro-ph/0003483 [astro-ph]].

\bibitem{Yoshida:2000uw}
N.~Yoshida, V.~Springel, S.~D.~M.~White and G.~Tormen,
``Weakly self-interacting dark matter and the structure of dark halos,''
Astrophys. J. Lett. \textbf{544}, L87-L90 (2000)
doi:10.1086/317306
[arXiv:astro-ph/0006134 [astro-ph]].

\bibitem{Dave:2000ar}
R.~Dave, D.~N.~Spergel, P.~J.~Steinhardt and B.~D.~Wandelt,
``Halo properties in cosmological simulations of selfinteracting cold dark matter,''
Astrophys. J. \textbf{547}, 574-589 (2001)
doi:10.1086/318417
[arXiv:astro-ph/0006218 [astro-ph]].

\bibitem{Colin:2002nk}
P.~Colin, V.~Avila-Reese, O.~Valenzuela and C.~Firmani,
``Structure and subhalo population of halos in a selfinteracting dark matter cosmology,''
Astrophys. J. \textbf{581}, 777-793 (2002)
doi:10.1086/344259
[arXiv:astro-ph/0205322 [astro-ph]].

\bibitem{Vogelsberger:2012ku}
M.~Vogelsberger, J.~Zavala and A.~Loeb,
``Subhaloes in Self-Interacting Galactic Dark Matter Haloes,''
Mon. Not. Roy. Astron. Soc. \textbf{423}, 3740 (2012)
doi:10.1111/j.1365-2966.2012.21182.x
[arXiv:1201.5892 [astro-ph.CO]].

\bibitem{Zavala:2012us}
J.~Zavala, M.~Vogelsberger and M.~G.~Walker,
``Constraining Self-Interacting Dark Matter with the Milky Way's dwarf spheroidals,''
Mon. Not. Roy. Astron. Soc. \textbf{431}, L20-L24 (2013)
doi:10.1093/mnrasl/sls053
[arXiv:1211.6426 [astro-ph.CO]].

\bibitem{Elbert:2014bma}
O.~D.~Elbert, J.~S.~Bullock, S.~Garrison-Kimmel, M.~Rocha, J.~O\~norbe and A.~H.~G.~Peter,
``Core formation in dwarf haloes with self-interacting dark matter: no fine-tuning necessary,''
Mon. Not. Roy. Astron. Soc. \textbf{453}, no.1, 29-37 (2015)
doi:10.1093/mnras/stv1470
[arXiv:1412.1477 [astro-ph.GA]].

\bibitem{Vogelsberger:2014pda}
M.~Vogelsberger, J.~Zavala, C.~Simpson and A.~Jenkins,
``Dwarf galaxies in CDM and SIDM with baryons: observational probes of the nature of dark matter,''
Mon. Not. Roy. Astron. Soc. \textbf{444}, no.4, 3684-3698 (2014)
doi:10.1093/mnras/stu1713
[arXiv:1405.5216 [astro-ph.CO]].

\bibitem{Fry:2015rta}
A.~B.~Fry, F.~Governato, A.~Pontzen, T.~Quinn, M.~Tremmel, L.~Anderson, H.~Menon, A.~M.~Brooks and J.~Wadsley,
``All about baryons: revisiting SIDM predictions at small halo masses,''
Mon. Not. Roy. Astron. Soc. \textbf{452}, no.2, 1468-1479 (2015)
doi:10.1093/mnras/stv1330
[arXiv:1501.00497 [astro-ph.CO]].

\bibitem{Dooley:2016ajo}
G.~A.~Dooley, A.~H.~G.~Peter, M.~Vogelsberger, J.~Zavala and A.~Frebel,
``Enhanced Tidal Stripping of Satellites in the Galactic Halo from Dark Matter Self-Interactions,''
Mon. Not. Roy. Astron. Soc. \textbf{461}, no.1, 710-727 (2016)
doi:10.1093/mnras/stw1309
[arXiv:1603.08919 [astro-ph.GA]].

\bibitem{Harvey:2015hha}
D.~Harvey, R.~Massey, T.~Kitching, A.~Taylor and E.~Tittley,
``The non-gravitational interactions of dark matter in colliding galaxy clusters,''
Science \textbf{347}, 1462-1465 (2015)
doi:10.1126/science.1261381
[arXiv:1503.07675 [astro-ph.CO]].

\bibitem{Kaplinghat:2015aga}
M.~Kaplinghat, S.~Tulin and H.~B.~Yu,
``Dark Matter Halos as Particle Colliders: Unified Solution to Small-Scale Structure Puzzles from Dwarfs to Clusters,''
Phys. Rev. Lett. \textbf{116}, no.4, 041302 (2016)
doi:10.1103/PhysRevLett.116.041302
[arXiv:1508.03339 [astro-ph.CO]].

\bibitem{Robertson:2018anx}
A.~Robertson, D.~Harvey, R.~Massey, V.~Eke, I.~G.~McCarthy, M.~Jauzac, B.~Li and J.~Schaye,
``Observable tests of self-interacting dark matter in galaxy clusters: cosmological simulations with SIDM and baryons,''
Mon. Not. Roy. Astron. Soc. \textbf{488}, no.3, 3646-3662 (2019)
doi:10.1093/mnras/stz1815
[arXiv:1810.05649 [astro-ph.CO]].

\bibitem{Sagunski:2020spe}
L.~Sagunski, S.~Gad-Nasr, B.~Colquhoun, A.~Robertson and S.~Tulin,
``Velocity-dependent Self-interacting Dark Matter from Groups and Clusters of Galaxies,''
JCAP \textbf{01}, 024 (2021)
doi:10.1088/1475-7516/2021/01/024
[arXiv:2006.12515 [astro-ph.CO]].

\bibitem{Andrade:2020lqq}
K.~E.~Andrade, J.~Fuson, S.~Gad-Nasr, D.~Kong, Q.~Minor, M.~G.~Roberts and M.~Kaplinghat,
``A stringent upper limit on dark matter self-interaction cross-section from cluster strong lensing,''
Mon. Not. Roy. Astron. Soc. \textbf{510}, no.1, 54-81 (2021)
doi:10.1093/mnras/stab3241
[arXiv:2012.06611 [astro-ph.CO]].

\bibitem{Elbert:2016dbb}
O.~D.~Elbert, J.~S.~Bullock, M.~Kaplinghat, S.~Garrison-Kimmel, A.~S.~Graus and M.~Rocha,
``A Testable Conspiracy: Simulating Baryonic Effects on Self-Interacting Dark Matter Halos,''
Astrophys. J. \textbf{853}, no.2, 109 (2018)
doi:10.3847/1538-4357/aa9710
[arXiv:1609.08626 [astro-ph.GA]].



\bibitem{Tulin:2017ara}
S.~Tulin and H.~B.~Yu,
``Dark Matter Self-interactions and Small Scale Structure,''
Phys. Rept. \textbf{730}, 1-57 (2018)
doi:10.1016/j.physrep.2017.11.004
[arXiv:1705.02358 [hep-ph]].



\bibitem{Buckley:2009in}
M.~R.~Buckley and P.~J.~Fox,
``Dark Matter Self-Interactions and Light Force Carriers,''
Phys. Rev. D \textbf{81}, 083522 (2010)
doi:10.1103/PhysRevD.81.083522
[arXiv:0911.3898 [hep-ph]].

\bibitem{Feng:2009hw}
J.~L.~Feng, M.~Kaplinghat and H.~B.~Yu,
``Halo Shape and Relic Density Exclusions of Sommerfeld-Enhanced Dark Matter Explanations of Cosmic Ray Excesses,''
Phys. Rev. Lett. \textbf{104}, 151301 (2010)
doi:10.1103/PhysRevLett.104.151301
[arXiv:0911.0422 [hep-ph]].

\bibitem{Feng:2009mn}
J.~L.~Feng, M.~Kaplinghat, H.~Tu and H.~B.~Yu,
``Hidden Charged Dark Matter,''
JCAP \textbf{07}, 004 (2009)
doi:10.1088/1475-7516/2009/07/004
[arXiv:0905.3039 [hep-ph]].

\bibitem{Loeb:2010gj}
A.~Loeb and N.~Weiner,
``Cores in Dwarf Galaxies from Dark Matter with a Yukawa Potential,''
Phys. Rev. Lett. \textbf{106}, 171302 (2011)
doi:10.1103/PhysRevLett.106.171302
[arXiv:1011.6374 [astro-ph.CO]].


\bibitem{vandenAarssen:2012vpm}
L.~G.~van den Aarssen, T.~Bringmann and C.~Pfrommer,
``Is dark matter with long-range interactions a solution to all small-scale problems of \textbackslash{}Lambda CDM cosmology?,''
Phys. Rev. Lett. \textbf{109}, 231301 (2012)
doi:10.1103/PhysRevLett.109.231301
[arXiv:1205.5809 [astro-ph.CO]].

\bibitem{Tulin:2013teo}
S.~Tulin, H.~B.~Yu and K.~M.~Zurek,
``Beyond Collisionless Dark Matter: Particle Physics Dynamics for Dark Matter Halo Structure,''
Phys. Rev. D \textbf{87}, no.11, 115007 (2013)
doi:10.1103/PhysRevD.87.115007
[arXiv:1302.3898 [hep-ph]].

\bibitem{Schutz:2014nka}
K.~Schutz and T.~R.~Slatyer,
``Self-Scattering for Dark Matter with an Excited State,''
JCAP \textbf{01}, 021 (2015)
doi:10.1088/1475-7516/2015/01/021
[arXiv:1409.2867 [hep-ph]].

\bibitem{Tulin:2012wi}
S.~Tulin, H.~B.~Yu and K.~M.~Zurek,
``Resonant Dark Forces and Small Scale Structure,''
Phys. Rev. Lett. \textbf{110}, no.11, 111301 (2013)
doi:10.1103/PhysRevLett.110.111301
[arXiv:1210.0900 [hep-ph]].

\bibitem{Boddy:2014qxa}
K.~K.~Boddy, J.~L.~Feng, M.~Kaplinghat, Y.~Shadmi and T.~M.~P.~Tait,
``Strongly interacting dark matter: Self-interactions and keV lines,''
Phys. Rev. D \textbf{90}, no.9, 095016 (2014)
doi:10.1103/PhysRevD.90.095016
[arXiv:1408.6532 [hep-ph]].

\bibitem{Ko:2014nha}
P.~Ko and Y.~Tang,
``Self-interacting scalar dark matter with local $Z_3$ symmetry,''
JCAP \textbf{05}, 047 (2014)
doi:10.1088/1475-7516/2014/05/047
[arXiv:1402.6449 [hep-ph]].

\bibitem{Kang:2015aqa}
Z.~Kang,
``View FImP miracle (by scale invariance) \`a la self-interaction,''
Phys. Lett. B \textbf{751}, 201-204 (2015)
doi:10.1016/j.physletb.2015.10.031
[arXiv:1505.06554 [hep-ph]].

\bibitem{Kainulainen:2015sva}
K.~Kainulainen, K.~Tuominen and V.~Vaskonen,
``Self-interacting dark matter and cosmology of a light scalar mediator,''
Phys. Rev. D \textbf{93}, no.1, 015016 (2016)
[erratum: Phys. Rev. D \textbf{95}, no.7, 079901 (2017)]
doi:10.1103/PhysRevD.93.015016
[arXiv:1507.04931 [hep-ph]].

\bibitem{Wang:2016lvj}
W.~Wang, M.~Zhang and J.~Zhao,
``Higgs exotic decays in general NMSSM with self-interacting dark matter,''
Int. J. Mod. Phys. A \textbf{33}, no.11, 1841002 (2018)
doi:10.1142/S0217751X18410026
[arXiv:1604.00123 [hep-ph]].

\bibitem{Duerr:2018mbd}
M.~Duerr, K.~Schmidt-Hoberg and S.~Wild,
``Self-interacting dark matter with a stable vector mediator,''
JCAP \textbf{09}, 033 (2018)
doi:10.1088/1475-7516/2018/09/033
[arXiv:1804.10385 [hep-ph]].

\bibitem{Kitahara:2016zyb}
T.~Kitahara and Y.~Yamamoto,
``Protophobic Light Vector Boson as a Mediator to the Dark Sector,''
Phys. Rev. D \textbf{95}, no.1, 015008 (2017)
doi:10.1103/PhysRevD.95.015008
[arXiv:1609.01605 [hep-ph]].

\bibitem{Ma:2017ucp}
E.~Ma,
``Inception of Self-Interacting Dark Matter with Dark Charge Conjugation Symmetry,''
Phys. Lett. B \textbf{772}, 442-445 (2017)
doi:10.1016/j.physletb.2017.06.067
[arXiv:1704.04666 [hep-ph]].

\bibitem{Bellazzini:2013foa}
B.~Bellazzini, M.~Cliche and P.~Tanedo,
``Effective theory of self-interacting dark matter,''
Phys. Rev. D \textbf{88}, no.8, 083506 (2013)
doi:10.1103/PhysRevD.88.083506
[arXiv:1307.1129 [hep-ph]].

\bibitem{Kamada:2020buc}
A.~Kamada, H.~J.~Kim and T.~Kuwahara,
``Maximally self-interacting dark matter: models and predictions,''
JHEP \textbf{12}, 202 (2020)
doi:10.1007/JHEP12(2020)202
[arXiv:2007.15522 [hep-ph]].

\bibitem{Kamada:2019jch}
A.~Kamada, M.~Yamada and T.~T.~Yanagida,
``Unification for darkly charged dark matter,''
Phys. Rev. D \textbf{102}, no.1, 015012 (2020)
doi:10.1103/PhysRevD.102.015012
[arXiv:1908.00207 [hep-ph]].

\bibitem{Kamada:2019gpp}
A.~Kamada, M.~Yamada and T.~T.~Yanagida,
``Unification of the standard model and dark matter sectors in [SU(5)$\times$U(1)]$^4$,''
JHEP \textbf{07}, 180 (2019)
doi:10.1007/JHEP07(2019)180
[arXiv:1905.04245 [hep-ph]].

\bibitem{Bringmann:2013vra}
T.~Bringmann, J.~Hasenkamp and J.~Kersten,
``Tight bonds between sterile neutrinos and dark matter,''
JCAP \textbf{07}, 042 (2014)
doi:10.1088/1475-7516/2014/07/042
[arXiv:1312.4947 [hep-ph]].

\bibitem{Ko:2014bka}
P.~Ko and Y.~Tang,
``\ensuremath{\nu}\ensuremath{\Lambda}MDM: A model for sterile neutrino and dark matter reconciles cosmological and neutrino oscillation data after BICEP2,''
Phys. Lett. B \textbf{739}, 62-67 (2014)
doi:10.1016/j.physletb.2014.10.035
[arXiv:1404.0236 [hep-ph]].

\bibitem{Kamada:2018zxi}
A.~Kamada, K.~Kaneta, K.~Yanagi and H.~B.~Yu,
``Self-interacting dark matter and muon $g-2$ in a gauged U$(1)_{L_{\mu} - L_{\tau}}$ model,''
JHEP \textbf{06}, 117 (2018)
doi:10.1007/JHEP06(2018)117
[arXiv:1805.00651 [hep-ph]].

\bibitem{Kamada:2018kmi}
A.~Kamada, M.~Yamada and T.~T.~Yanagida,
``Self-interacting dark matter with a vector mediator: kinetic mixing with the $ \mathrm{U}{(1)}_{{\left(B-L\right)}_3} $ gauge boson,''
JHEP \textbf{03}, 021 (2019)
doi:10.1007/JHEP03(2019)021
[arXiv:1811.02567 [hep-ph]].

\bibitem{Aboubrahim:2020lnr}
A.~Aboubrahim, W.~Z.~Feng, P.~Nath and Z.~Y.~Wang,
``Self-interacting hidden sector dark matter, small scale galaxy structure anomalies, and a dark force,''
Phys. Rev. D \textbf{103}, no.7, 075014 (2021)
doi:10.1103/PhysRevD.103.075014
[arXiv:2008.00529 [hep-ph]].



\bibitem{Pospelov:2007mp}
M.~Pospelov, A.~Ritz and M.~B.~Voloshin,
``Secluded WIMP Dark Matter,''
Phys. Lett. B \textbf{662}, 53-61 (2008)
doi:10.1016/j.physletb.2008.02.052
[arXiv:0711.4866 [hep-ph]].



\bibitem{DirectSearch1} 
  A.~Tan {\it et al.} [PandaX-II Collaboration],
 ``Dark Matter Results from First 98.7 Days of Data from the PandaX-II Experiment,''
  Phys.\ Rev.\ Lett.\  {\bf 117}, no. 12, 121303 (2016)
  doi:10.1103/PhysRevLett.117.121303
  [arXiv:1607.07400 [hep-ex]].
  \bibitem{DirectSearch2} 
  D.~S.~Akerib {\it et al.} [LUX Collaboration],
``Results from a search for dark matter in the complete LUX exposure,''
  Phys.\ Rev.\ Lett.\  {\bf 118}, no. 2, 021303 (2017)
  doi:10.1103/PhysRevLett.118.021303
  [arXiv:1608.07648 [astro-ph.CO]].


\bibitem{LHCSearch1}   
  V.~Khachatryan {\it et al.} [CMS Collaboration],
  ``Search for dark matter particles in proton-proton collisions at $ \sqrt{s}=8 $ TeV using the razor variables,''
  JHEP {\bf 1612}, 088 (2016)
  doi:10.1007/JHEP12(2016)088
  [arXiv:1603.08914 [hep-ex]].
  \bibitem{LHCSearch2}   
  G.~Aad {\it et al.} [ATLAS Collaboration],
  ``Search for dark matter in events with a hadronically decaying W or Z boson and missing transverse momentum in $pp$ collisions at $\sqrt{s} =$ 8 TeV with the ATLAS detector,''
  Phys.\ Rev.\ Lett.\  {\bf 112}, no. 4, 041802 (2014)
  doi:10.1103/PhysRevLett.112.041802
  [arXiv:1309.4017 [hep-ex]].
  \bibitem{LHCSearch3}   
  D.~Abercrombie {\it et al.},
  ``Dark Matter Benchmark Models for Early LHC Run-2 Searches: Report of the ATLAS/CMS Dark Matter Forum,''
  arXiv:1507.00966 [hep-ex].
  \bibitem{LHCSearch4}   
  M.~Aaboud {\it et al.} [ATLAS Collaboration],
  ``Search for dark matter at $\sqrt{s}=13$ TeV in final states containing an energetic photon and large missing transverse momentum with the ATLAS detector,''
  Eur.\ Phys.\ J.\ C {\bf 77}, no. 6, 393 (2017)
  doi:10.1140/epjc/s10052-017-4965-8
  [arXiv:1704.03848 [hep-ex]].
  \bibitem{LHCSearch5}   
  M.~Aaboud {\it et al.} [ATLAS Collaboration],
  ``Search for Dark Matter Produced in Association with a Higgs Boson Decaying to $b\bar b$ using 36 fb$^{-1}$ of $pp$ collisions at $\sqrt s=13$ TeV with the ATLAS Detector,''
  Phys.\ Rev.\ Lett.\  {\bf 119}, no. 18, 181804 (2017)
  doi:10.1103/PhysRevLett.119.181804
  [arXiv:1707.01302 [hep-ex]].





\bibitem{Kamionkowski:2008gj}
M.~Kamionkowski and S.~Profumo,
``Early Annihilation and Diffuse Backgrounds in Models of Weakly Interacting Massive Particles in Which the Cross Section for Pair Annihilation Is Enhanced by 1/v,''
Phys. Rev. Lett. \textbf{101}, 261301 (2008)
doi:10.1103/PhysRevLett.101.261301
[arXiv:0810.3233 [astro-ph]].

\bibitem{Zavala:2009mi}
J.~Zavala, M.~Vogelsberger and S.~D.~M.~White,
``Relic density and CMB constraints on dark matter annihilation with Sommerfeld enhancement,''
Phys. Rev. D \textbf{81}, 083502 (2010)
doi:10.1103/PhysRevD.81.083502
[arXiv:0910.5221 [astro-ph.CO]].

\bibitem{Feng:2010zp}
J.~L.~Feng, M.~Kaplinghat and H.~B.~Yu,
``Sommerfeld Enhancements for Thermal Relic Dark Matter,''
Phys. Rev. D \textbf{82}, 083525 (2010)
doi:10.1103/PhysRevD.82.083525
[arXiv:1005.4678 [hep-ph]].

\bibitem{Hisano:2011dc}
J.~Hisano, M.~Kawasaki, K.~Kohri, T.~Moroi, K.~Nakayama and T.~Sekiguchi,
``Cosmological constraints on dark matter models with velocity-dependent annihilation cross section,''
Phys. Rev. D \textbf{83}, 123511 (2011)
doi:10.1103/PhysRevD.83.123511
[arXiv:1102.4658 [hep-ph]].

\bibitem{Bergstrom:2008ag}
L.~Bergstr\"om, G.~Bertone, T.~Bringmann, J.~Edsj\"o and M.~Taoso,
``Gamma-ray and Radio Constraints of High Positron Rate Dark Matter Models Annihilating into New Light Particles,''
Phys. Rev. D \textbf{79}, 081303 (2009)
doi:10.1103/PhysRevD.79.081303
[arXiv:0812.3895 [astro-ph]].

\bibitem{Mardon:2009rc}
J.~Mardon, Y.~Nomura, D.~Stolarski and J.~Thaler,
``Dark Matter Signals from Cascade Annihilations,''
JCAP \textbf{05}, 016 (2009)
doi:10.1088/1475-7516/2009/05/016
[arXiv:0901.2926 [hep-ph]].

\bibitem{Galli:2009zc}
S.~Galli, F.~Iocco, G.~Bertone and A.~Melchiorri,
``CMB constraints on Dark Matter models with large annihilation cross-section,''
Phys. Rev. D \textbf{80}, 023505 (2009)
doi:10.1103/PhysRevD.80.023505
[arXiv:0905.0003 [astro-ph.CO]].

\bibitem{Slatyer:2009yq}
T.~R.~Slatyer, N.~Padmanabhan and D.~P.~Finkbeiner,
``CMB Constraints on WIMP Annihilation: Energy Absorption During the Recombination Epoch,''
Phys. Rev. D \textbf{80}, 043526 (2009)
doi:10.1103/PhysRevD.80.043526
[arXiv:0906.1197 [astro-ph.CO]].

\bibitem{Hannestad:2010zt}
S.~Hannestad and T.~Tram,
``Sommerfeld Enhancement of DM Annihilation: Resonance Structure, Freeze-Out and CMB Spectral Bound,''
JCAP \textbf{01}, 016 (2011)
doi:10.1088/1475-7516/2011/01/016
[arXiv:1008.1511 [astro-ph.CO]].

\bibitem{Finkbeiner:2010sm}
D.~P.~Finkbeiner, L.~Goodenough, T.~R.~Slatyer, M.~Vogelsberger and N.~Weiner,
``Consistent Scenarios for Cosmic-Ray Excesses from Sommerfeld-Enhanced Dark Matter Annihilation,''
JCAP \textbf{05}, 002 (2011)
doi:10.1088/1475-7516/2011/05/002
[arXiv:1011.3082 [hep-ph]].





\bibitem{Sommerfeld}
A. Sommerfeld, Annalen der Physik 403, 257 (1931).

\bibitem{Hisano:2003ec}
J.~Hisano, S.~Matsumoto and M.~M.~Nojiri,
``Explosive dark matter annihilation,''
Phys. Rev. Lett. \textbf{92}, 031303 (2004)
doi:10.1103/PhysRevLett.92.031303
[arXiv:hep-ph/0307216 [hep-ph]].

\bibitem{Hisano:2004ds}
J.~Hisano, S.~Matsumoto, M.~M.~Nojiri and O.~Saito,
``Non-perturbative effect on dark matter annihilation and gamma ray signature from galactic center,''
Phys. Rev. D \textbf{71}, 063528 (2005)
doi:10.1103/PhysRevD.71.063528
[arXiv:hep-ph/0412403 [hep-ph]].

\bibitem{Hisano:2005ec}
J.~Hisano, S.~Matsumoto, O.~Saito and M.~Senami,
``Heavy wino-like neutralino dark matter annihilation into antiparticles,''
Phys. Rev. D \textbf{73}, 055004 (2006)
doi:10.1103/PhysRevD.73.055004
[arXiv:hep-ph/0511118 [hep-ph]].

\bibitem{Cirelli:2007xd}
M.~Cirelli, A.~Strumia and M.~Tamburini,
``Cosmology and Astrophysics of Minimal Dark Matter,''
Nucl. Phys. B \textbf{787}, 152-175 (2007)
doi:10.1016/j.nuclphysb.2007.07.023
[arXiv:0706.4071 [hep-ph]].

\bibitem{Arkani-Hamed:2008hhe}
N.~Arkani-Hamed, D.~P.~Finkbeiner, T.~R.~Slatyer and N.~Weiner,
``A Theory of Dark Matter,''
Phys. Rev. D \textbf{79}, 015014 (2009)
doi:10.1103/PhysRevD.79.015014
[arXiv:0810.0713 [hep-ph]].

\bibitem{Cholis:2008qq}
I.~Cholis, D.~P.~Finkbeiner, L.~Goodenough and N.~Weiner,
``The PAMELA Positron Excess from Annihilations into a Light Boson,''
JCAP \textbf{12}, 007 (2009)
doi:10.1088/1475-7516/2009/12/007
[arXiv:0810.5344 [astro-ph]].






\bibitem{Bringmann:2016din}
T.~Bringmann, F.~Kahlhoefer, K.~Schmidt-Hoberg and P.~Walia,
``Strong constraints on self-interacting dark matter with light mediators,''
Phys. Rev. Lett. \textbf{118}, no.14, 141802 (2017)
doi:10.1103/PhysRevLett.118.141802
[arXiv:1612.00845 [hep-ph]].





\bibitem{Kaplan:2009ag} 
  D.~E.~Kaplan, M.~A.~Luty and K.~M.~Zurek,
``Asymmetric Dark Matter,''
  Phys.\ Rev.\ D {\bf 79}, 115016 (2009)
  doi:10.1103/PhysRevD.79.115016
  [arXiv:0901.4117 [hep-ph]].
  
 \bibitem{Petraki:2013wwa} 
  K.~Petraki and R.~R.~Volkas,
``Review of asymmetric dark matter,''
  Int.\ J.\ Mod.\ Phys.\ A {\bf 28}, 1330028 (2013)
  doi:10.1142/S0217751X13300287
  [arXiv:1305.4939 [hep-ph]].
  
 \bibitem{Zurek:2013wia} 
  K.~M.~Zurek,
 ``Asymmetric Dark Matter: Theories, Signatures, and Constraints,''
  Phys.\ Rept.\  {\bf 537}, 91 (2014)
  doi:10.1016/j.physrep.2013.12.001
  [arXiv:1308.0338 [hep-ph]].
  



\bibitem{Lin:2011gj}
T.~Lin, H.~B.~Yu and K.~M.~Zurek,
``On Symmetric and Asymmetric Light Dark Matter,''
Phys. Rev. D \textbf{85}, 063503 (2012)
doi:10.1103/PhysRevD.85.063503
[arXiv:1111.0293 [hep-ph]].

\bibitem{Baldes:2017gzu}
I.~Baldes, M.~Cirelli, P.~Panci, K.~Petraki, F.~Sala and M.~Taoso,
``Asymmetric dark matter: residual annihilations and self-interactions,''
SciPost Phys. \textbf{4}, no.6, 041 (2018)
doi:10.21468/SciPostPhys.4.6.041
[arXiv:1712.07489 [hep-ph]].




\bibitem{Nussinov:1985xr} 
  S.~Nussinov,
``Technocosmology: Could A Technibaryon Excess Provide A 'natural' Missing Mass Candidate?,''
  Phys.\ Lett.\  {\bf 165B}, 55 (1985).
  doi:10.1016/0370-2693(85)90689-6

\bibitem{Kaplan:1991ah} 
  D.~B.~Kaplan,
``A Single explanation for both the baryon and dark matter densities,''
  Phys.\ Rev.\ Lett.\  {\bf 68}, 741 (1992).
  doi:10.1103/PhysRevLett.68.741

\bibitem{Barr:1990ca} 
  S.~M.~Barr, R.~S.~Chivukula and E.~Farhi,
 ``Electroweak Fermion Number Violation and the Production of Stable Particles in the Early Universe,''
  Phys.\ Lett.\ B {\bf 241}, 387 (1990).
  doi:10.1016/0370-2693(90)91661-T

\bibitem{Barr:1991qn} 
  S.~M.~Barr,
 ``Baryogenesis, sphalerons and the cogeneration of dark matter,''
  Phys.\ Rev.\ D {\bf 44}, 3062 (1991).
  doi:10.1103/PhysRevD.44.3062

\bibitem{Dodelson:1991iv} 
  S.~Dodelson, B.~R.~Greene and L.~M.~Widrow,
``Baryogenesis, dark matter and the width of the Z,''
  Nucl.\ Phys.\ B {\bf 372}, 467 (1992).
  doi:10.1016/0550-3213(92)90328-9

\bibitem{Fujii:2002aj} 
  M.~Fujii and T.~Yanagida,
``A Solution to the coincidence puzzle of Omega(B) and Omega (DM),''
  Phys.\ Lett.\ B {\bf 542}, 80 (2002)
  doi:10.1016/S0370-2693(02)02341-9
  [hep-ph/0206066].

\bibitem{Kitano:2004sv} 
  R.~Kitano and I.~Low,
``Dark matter from baryon asymmetry,''
  Phys.\ Rev.\ D {\bf 71}, 023510 (2005)
  doi:10.1103/PhysRevD.71.023510
  [hep-ph/0411133].

\bibitem{Farrar:2005zd} 
  G.~R.~Farrar and G.~Zaharijas,
``Dark matter and the baryon asymmetry,''
  Phys.\ Rev.\ Lett.\  {\bf 96}, 041302 (2006)
  doi:10.1103/PhysRevLett.96.041302
  [hep-ph/0510079].

\bibitem{Kitano:2008tk} 
  R.~Kitano, H.~Murayama and M.~Ratz,
``Unified origin of baryons and dark matter,''
  Phys.\ Lett.\ B {\bf 669}, 145 (2008)
  doi:10.1016/j.physletb.2008.09.049
  [arXiv:0807.4313 [hep-ph]].

\bibitem{Gudnason:2006ug} 
  S.~B.~Gudnason, C.~Kouvaris and F.~Sannino,
``Towards working technicolor: Effective theories and dark matter,''
  Phys.\ Rev.\ D {\bf 73}, 115003 (2006)
  doi:10.1103/PhysRevD.73.115003
  [hep-ph/0603014].


  
 \bibitem{Shelton:2010ta} 
  J.~Shelton and K.~M.~Zurek,
``Darkogenesis: A baryon asymmetry from the dark matter sector,''
  Phys.\ Rev.\ D {\bf 82}, 123512 (2010)
  doi:10.1103/PhysRevD.82.123512
  [arXiv:1008.1997 [hep-ph]].
  
 \bibitem{Davoudiasl:2010am} 
  H.~Davoudiasl, D.~E.~Morrissey, K.~Sigurdson and S.~Tulin,
``Hylogenesis: A Unified Origin for Baryonic Visible Matter and Antibaryonic Dark Matter,''
  Phys.\ Rev.\ Lett.\  {\bf 105}, 211304 (2010)
  doi:10.1103/PhysRevLett.105.211304
  [arXiv:1008.2399 [hep-ph]].
  
\bibitem{Huang:2017kzu}
F.~P.~Huang and C.~S.~Li,
``Probing the baryogenesis and dark matter relaxed in phase transition by gravitational waves and colliders,''
Phys. Rev. D \textbf{96}, no.9, 095028 (2017)
doi:10.1103/PhysRevD.96.095028
[arXiv:1709.09691 [hep-ph]].
  
  \bibitem{Buckley:2010ui} 
  M.~R.~Buckley and L.~Randall,
``Xogenesis,''
  JHEP {\bf 1109}, 009 (2011)
  doi:10.1007/JHEP09(2011)009
  [arXiv:1009.0270 [hep-ph]].
  
  \bibitem{Cohen:2010kn} 
  T.~Cohen, D.~J.~Phalen, A.~Pierce and K.~M.~Zurek,
``Asymmetric Dark Matter from a GeV Hidden Sector,''
  Phys.\ Rev.\ D {\bf 82}, 056001 (2010)
  doi:10.1103/PhysRevD.82.056001
  [arXiv:1005.1655 [hep-ph]].
    
  \bibitem{Ibe:2018juk} 
  M.~Ibe, A.~Kamada, S.~Kobayashi and W.~Nakano,
  ``Composite Asymmetric Dark Matter with a Dark Photon Portal,''
  JHEP {\bf 1811}, 203 (2018)
  doi:10.1007/JHEP11(2018)203
  [arXiv:1805.06876 [hep-ph]].
  
  \bibitem{Ibe:2018tex} 
  M.~Ibe, A.~Kamada, S.~Kobayashi, T.~Kuwahara and W.~Nakano,
  ``Ultraviolet Completion of a Composite Asymmetric Dark Matter Model with a Dark Photon Portal,''
  JHEP {\bf 1903}, 173 (2019)
  doi:10.1007/JHEP03(2019)173
  [arXiv:1811.10232 [hep-ph]].

\bibitem{An:2009vq}
H.~An, S.~L.~Chen, R.~N.~Mohapatra and Y.~Zhang,
``Leptogenesis as a Common Origin for Matter and Dark Matter,''
JHEP \textbf{03}, 124 (2010)
doi:10.1007/JHEP03(2010)124
[arXiv:0911.4463 [hep-ph]].

\bibitem{Falkowski:2011xh}
A.~Falkowski, J.~T.~Ruderman and T.~Volansky,
``Asymmetric Dark Matter from Leptogenesis,''
JHEP \textbf{05}, 106 (2011)
doi:10.1007/JHEP05(2011)106
[arXiv:1101.4936 [hep-ph]].

\bibitem{Bai:2013xga}
Y.~Bai and P.~Schwaller,
``Scale of dark QCD,''
Phys. Rev. D \textbf{89}, no.6, 063522 (2014)
doi:10.1103/PhysRevD.89.063522
[arXiv:1306.4676 [hep-ph]].

\bibitem{Zhang:2021orr}
M.~Zhang,
``Leptophilic composite asymmetric dark matter and its detection,''
Phys. Rev. D \textbf{104}, no.5, 055008 (2021)
doi:10.1103/PhysRevD.104.055008
[arXiv:2104.06988 [hep-ph]].

\bibitem{Alves:2009nf}
D.~S.~M.~Alves, S.~R.~Behbahani, P.~Schuster and J.~G.~Wacker,
``Composite Inelastic Dark Matter,''
Phys. Lett. B \textbf{692}, 323-326 (2010)
doi:10.1016/j.physletb.2010.08.006
[arXiv:0903.3945 [hep-ph]].

\bibitem{Alves:2010dd}
D.~Spier Moreira Alves, S.~R.~Behbahani, P.~Schuster and J.~G.~Wacker,
``The Cosmology of Composite Inelastic Dark Matter,''
JHEP \textbf{06}, 113 (2010)
doi:10.1007/JHEP06(2010)113
[arXiv:1003.4729 [hep-ph]].

\bibitem{Beylin:2020bsz}
V.~Beylin, M.~Khlopov, V.~Kuksa and N.~Volchanskiy,
``New physics of strong interaction and Dark Universe,''
Universe \textbf{6}, no.11, 196 (2020)
doi:10.3390/universe6110196
[arXiv:2010.13678 [hep-ph]].

\bibitem{Khlopov:1989fj}
M.~Y.~Khlopov, G.~M.~Beskin, N.~E.~Bochkarev, L.~A.~Pustylnik and S.~A.~Pustylnik,
``Observational Physics of Mirror World,''
Sov. Astron. \textbf{35}, 21 (1991)
FERMILAB-PUB-89-193-A.

\bibitem{Blinnikov:1983gh}
S.~I.~Blinnikov and M.~Khlopov,
``Possible astronomical effects of mirror particles,''
Sov. Astron. \textbf{27}, 371-375 (1983)

\bibitem{Blinnikov:1982eh}
S.~I.~Blinnikov and M.~Y.~Khlopov,
``ON POSSIBLE EFFECTS OF 'MIRROR' PARTICLES,''
Sov. J. Nucl. Phys. \textbf{36}, 472 (1982)
ITEP-11-1982.


\bibitem{Blennow:2012de}
M.~Blennow, E.~Fernandez-Martinez, O.~Mena, J.~Redondo and P.~Serra,
``Asymmetric Dark Matter and Dark Radiation,''
JCAP \textbf{07}, 022 (2012)
doi:10.1088/1475-7516/2012/07/022
[arXiv:1203.5803 [hep-ph]].

\bibitem{Frandsen:2011kt}
M.~T.~Frandsen, S.~Sarkar and K.~Schmidt-Hoberg,
``Light asymmetric dark matter from new strong dynamics,''
Phys. Rev. D \textbf{84}, 051703 (2011)
doi:10.1103/PhysRevD.84.051703
[arXiv:1103.4350 [hep-ph]].

\bibitem{Murgui:2021eqf}
C.~Murgui and K.~M.~Zurek,
``Dark Unification: a UV-complete Theory of Asymmetric Dark Matter,''
[arXiv:2112.08374 [hep-ph]].

\bibitem{Kamada:2021cow}
A.~Kamada and T.~Kuwahara,
``LHC lifetime frontier and visible decay searches in composite asymmetric dark matter models,''
JHEP \textbf{03}, 176 (2022)
doi:10.1007/JHEP03(2022)176
[arXiv:2112.01202 [hep-ph]].

\bibitem{Ibe:2019ena}
M.~Ibe, A.~Kamada, S.~Kobayashi, T.~Kuwahara and W.~Nakano,
``Baryon-Dark Matter Coincidence in Mirrored Unification,''
Phys. Rev. D \textbf{100}, no.7, 075022 (2019)
doi:10.1103/PhysRevD.100.075022
[arXiv:1907.03404 [hep-ph]].




\bibitem{Mohapatra:2001sx}
R.~N.~Mohapatra, S.~Nussinov and V.~L.~Teplitz,
``Mirror matter as selfinteracting dark matter,''
Phys. Rev. D \textbf{66}, 063002 (2002)
doi:10.1103/PhysRevD.66.063002
[arXiv:hep-ph/0111381 [hep-ph]].


\bibitem{Frandsen:2010yj}
M.~T.~Frandsen and S.~Sarkar,
``Asymmetric dark matter and the Sun,''
Phys. Rev. Lett. \textbf{105}, 011301 (2010)
doi:10.1103/PhysRevLett.105.011301
[arXiv:1003.4505 [hep-ph]].

\bibitem{Petraki:2014uza}
K.~Petraki, L.~Pearce and A.~Kusenko,
``Self-interacting asymmetric dark matter coupled to a light massive dark photon,''
JCAP \textbf{07}, 039 (2014)
doi:10.1088/1475-7516/2014/07/039
[arXiv:1403.1077 [hep-ph]].

\bibitem{Dessert:2018khu}
C.~Dessert, C.~Kilic, C.~Trendafilova and Y.~Tsai,
``Addressing Astrophysical and Cosmological Problems With Secretly Asymmetric Dark Matter,''
Phys. Rev. D \textbf{100}, no.1, 015029 (2019)
doi:10.1103/PhysRevD.100.015029
[arXiv:1811.05534 [hep-ph]].

\bibitem{Dutta:2022knf}
M.~Dutta, N.~Narendra, N.~Sahu and S.~Shil,
``Asymmetric Self-interacting Dark Matter via Dirac Leptogenesis,''
[arXiv:2202.04704 [hep-ph]].

\bibitem{Heeck:2022znj}
J.~Heeck and A.~Thapa,
``Explaining lepton-flavor non-universality and self-interacting dark matter with $L_\mu-L_\tau$,''
[arXiv:2202.08854 [hep-ph]].









\bibitem{Perez:2021udy}
P.~F.~Perez, C.~Murgui and A.~D.~Plascencia,
``Baryogenesis via leptogenesis: Spontaneous B and L violation,''
Phys. Rev. D \textbf{104}, no.5, 055007 (2021)
doi:10.1103/PhysRevD.104.055007
[arXiv:2103.13397 [hep-ph]].






\bibitem{Buckley:2011ye}
M.~R.~Buckley and S.~Profumo,
``Regenerating a Symmetry in Asymmetric Dark Matter,''
Phys. Rev. Lett. \textbf{108}, 011301 (2012)
doi:10.1103/PhysRevLett.108.011301
[arXiv:1109.2164 [hep-ph]].

\bibitem{Tulin:2012re}
S.~Tulin, H.~B.~Yu and K.~M.~Zurek,
``Oscillating Asymmetric Dark Matter,''
JCAP \textbf{05}, 013 (2012)
doi:10.1088/1475-7516/2012/05/013
[arXiv:1202.0283 [hep-ph]].




\bibitem{Husdal:2016haj}
L.~Husdal,
``On Effective Degrees of Freedom in the Early Universe,''
Galaxies \textbf{4}, no.4, 78 (2016)
doi:10.3390/galaxies4040078
[arXiv:1609.04979 [astro-ph.CO]].



\bibitem{Fukugita:1986hr}
M.~Fukugita and T.~Yanagida,
``Baryogenesis Without Grand Unification,''
Phys. Lett. B \textbf{174}, 45-47 (1986)
doi:10.1016/0370-2693(86)91126-3

\bibitem{Buchmuller:2004nz}
W.~Buchmuller, P.~Di Bari and M.~Plumacher,
``Leptogenesis for pedestrians,''
Annals Phys. \textbf{315}, 305-351 (2005)
doi:10.1016/j.aop.2004.02.003
[arXiv:hep-ph/0401240 [hep-ph]].

\bibitem{Davidson:2008bu}
S.~Davidson, E.~Nardi and Y.~Nir,
``Leptogenesis,''
Phys. Rept. \textbf{466}, 105-177 (2008)
doi:10.1016/j.physrep.2008.06.002
[arXiv:0802.2962 [hep-ph]].

\bibitem{Covi:1996wh}
L.~Covi, E.~Roulet and F.~Vissani,
``CP violating decays in leptogenesis scenarios,''
Phys. Lett. B \textbf{384}, 169-174 (1996)
doi:10.1016/0370-2693(96)00817-9
[arXiv:hep-ph/9605319 [hep-ph]].





\bibitem{Scherrer:1985zt}
R.~J.~Scherrer and M.~S.~Turner,
``On the Relic, Cosmic Abundance of Stable Weakly Interacting Massive Particles,''
Phys. Rev. D \textbf{33}, 1585 (1986)
[erratum: Phys. Rev. D \textbf{34}, 3263 (1986)]
doi:10.1103/PhysRevD.33.1585

\bibitem{Griest:1986yu}
K.~Griest and D.~Seckel,
``Cosmic Asymmetry, Neutrinos and the Sun,''
Nucl. Phys. B \textbf{283}, 681-705 (1987)
[erratum: Nucl. Phys. B \textbf{296}, 1034-1036 (1988)]
doi:10.1016/0550-3213(87)90293-8

\bibitem{Graesser:2011wi}
M.~L.~Graesser, I.~M.~Shoemaker and L.~Vecchi,
``Asymmetric WIMP dark matter,''
JHEP \textbf{10}, 110 (2011)
doi:10.1007/JHEP10(2011)110
[arXiv:1103.2771 [hep-ph]].

\bibitem{Iminniyaz:2011yp}
H.~Iminniyaz, M.~Drees and X.~Chen,
``Relic Abundance of Asymmetric Dark Matter,''
JCAP \textbf{07}, 003 (2011)
doi:10.1088/1475-7516/2011/07/003
[arXiv:1104.5548 [hep-ph]].

\bibitem{Bell:2014xta}
N.~F.~Bell, S.~Horiuchi and I.~M.~Shoemaker,
``Annihilating Asymmetric Dark Matter,''
Phys. Rev. D \textbf{91}, no.2, 023505 (2015)
doi:10.1103/PhysRevD.91.023505
[arXiv:1408.5142 [hep-ph]].

\bibitem{Murase:2016nwx}
K.~Murase and I.~M.~Shoemaker,
``Detecting Asymmetric Dark Matter in the Sun with Neutrinos,''
Phys. Rev. D \textbf{94}, no.6, 063512 (2016)
doi:10.1103/PhysRevD.94.063512
[arXiv:1606.03087 [hep-ph]].


\bibitem{Baldes:2017gzw}
I.~Baldes and K.~Petraki,
``Asymmetric thermal-relic dark matter: Sommerfeld-enhanced freeze-out, annihilation signals and unitarity bounds,''
JCAP \textbf{09}, 028 (2017)
doi:10.1088/1475-7516/2017/09/028
[arXiv:1703.00478 [hep-ph]].







\bibitem{Hufnagel:2018bjp}
M.~Hufnagel, K.~Schmidt-Hoberg and S.~Wild,
``BBN constraints on MeV-scale dark sectors. Part II. Electromagnetic decays,''
JCAP \textbf{11}, 032 (2018)
doi:10.1088/1475-7516/2018/11/032
[arXiv:1808.09324 [hep-ph]].

\bibitem{Depta:2020zbh}
P.~F.~Depta, M.~Hufnagel and K.~Schmidt-Hoberg,
``Updated BBN constraints on electromagnetic decays of MeV-scale particles,''
JCAP \textbf{04}, 011 (2021)
doi:10.1088/1475-7516/2021/04/011
[arXiv:2011.06519 [hep-ph]].

\bibitem{Ibe:2021fed}
M.~Ibe, S.~Kobayashi, Y.~Nakayama and S.~Shirai,
``Cosmological constraints on dark scalar,''
JHEP \textbf{03}, 198 (2022)
doi:10.1007/JHEP03(2022)198
[arXiv:2112.11096 [hep-ph]].




\bibitem{Cassel:2009wt}
S.~Cassel,
``Sommerfeld factor for arbitrary partial wave processes,''
J. Phys. G \textbf{37}, 105009 (2010)
doi:10.1088/0954-3899/37/10/105009
[arXiv:0903.5307 [hep-ph]].



\bibitem{Galli:2009zc}
S.~Galli, F.~Iocco, G.~Bertone and A.~Melchiorri,
``CMB constraints on Dark Matter models with large annihilation cross-section,''
Phys. Rev. D \textbf{80}, 023505 (2009)
doi:10.1103/PhysRevD.80.023505
[arXiv:0905.0003 [astro-ph.CO]].

\bibitem{Slatyer:2009yq}
T.~R.~Slatyer, N.~Padmanabhan and D.~P.~Finkbeiner,
``CMB Constraints on WIMP Annihilation: Energy Absorption During the Recombination Epoch,''
Phys. Rev. D \textbf{80}, 043526 (2009)
doi:10.1103/PhysRevD.80.043526
[arXiv:0906.1197 [astro-ph.CO]].

\bibitem{Cline:2013fm}
J.~M.~Cline and P.~Scott,
``Dark Matter CMB Constraints and Likelihoods for Poor Particle Physicists,''
JCAP \textbf{03}, 044 (2013)
[erratum: JCAP \textbf{05}, E01 (2013)]
doi:10.1088/1475-7516/2013/03/044
[arXiv:1301.5908 [astro-ph.CO]].

\bibitem{Liu:2016cnk}
H.~Liu, T.~R.~Slatyer and J.~Zavala,
``Contributions to cosmic reionization from dark matter annihilation and decay,''
Phys. Rev. D \textbf{94}, no.6, 063507 (2016)
doi:10.1103/PhysRevD.94.063507
[arXiv:1604.02457 [astro-ph.CO]].

\bibitem{Kawasaki:2021etm}
M.~Kawasaki, H.~Nakatsuka, K.~Nakayama and T.~Sekiguchi,
``Revisiting CMB constraints on dark matter annihilation,''
JCAP \textbf{12}, no.12, 015 (2021)
doi:10.1088/1475-7516/2021/12/015
[arXiv:2105.08334 [astro-ph.CO]].

\bibitem{Planck:2019nip}
N.~Aghanim \textit{et al.} [Planck],
``Planck 2018 results. V. CMB power spectra and likelihoods,''
Astron. Astrophys. \textbf{641}, A5 (2020)
doi:10.1051/0004-6361/201936386
[arXiv:1907.12875 [astro-ph.CO]].

\bibitem{BOSS:2013rlg}
L.~Anderson \textit{et al.} [BOSS],
``The clustering of galaxies in the SDSS-III Baryon Oscillation Spectroscopic Survey: baryon acoustic oscillations in the Data Releases 10 and 11 Galaxy samples,''
Mon. Not. Roy. Astron. Soc. \textbf{441}, no.1, 24-62 (2014)
doi:10.1093/mnras/stu523
[arXiv:1312.4877 [astro-ph.CO]].

\bibitem{Ross:2014qpa}
A.~J.~Ross, L.~Samushia, C.~Howlett, W.~J.~Percival, A.~Burden and M.~Manera,
``The clustering of the SDSS DR7 main Galaxy sample \textendash{} I. A 4 per cent distance measure at $z = 0.15$,''
Mon. Not. Roy. Astron. Soc. \textbf{449}, no.1, 835-847 (2015)
doi:10.1093/mnras/stv154
[arXiv:1409.3242 [astro-ph.CO]].

\bibitem{DES:2017myr}
T.~M.~C.~Abbott \textit{et al.} [DES],
``Dark Energy Survey year 1 results: Cosmological constraints from galaxy clustering and weak lensing,''
Phys. Rev. D \textbf{98}, no.4, 043526 (2018)
doi:10.1103/PhysRevD.98.043526
[arXiv:1708.01530 [astro-ph.CO]].





\bibitem{Planck:2018vyg}
N.~Aghanim \textit{et al.} [Planck],
``Planck 2018 results. VI. Cosmological parameters,''
Astron. Astrophys. \textbf{641}, A6 (2020)
[erratum: Astron. Astrophys. \textbf{652}, C4 (2021)]
doi:10.1051/0004-6361/201833910
[arXiv:1807.06209 [astro-ph.CO]].

\bibitem{deSalas:2016ztq}
P.~F.~de Salas and S.~Pastor,
``Relic neutrino decoupling with flavour oscillations revisited,''
JCAP \textbf{07}, 051 (2016)
doi:10.1088/1475-7516/2016/07/051
[arXiv:1606.06986 [hep-ph]].


\bibitem{Bai:2021ibt}
Y.~Bai and M.~Korwar,
``Cosmological constraints on first-order phase transitions,''
Phys. Rev. D \textbf{105}, no.9, 095015 (2022)
doi:10.1103/PhysRevD.105.095015
[arXiv:2109.14765 [hep-ph]].


\bibitem{CMB-S4:2022ght}
K.~Abazajian \textit{et al.} [CMB-S4],
``Snowmass 2021 CMB-S4 White Paper,''
[arXiv:2203.08024 [astro-ph.CO]].



\bibitem{Cyr-Racine:2013fsa}
F.~Y.~Cyr-Racine, R.~de Putter, A.~Raccanelli and K.~Sigurdson,
``Constraints on Large-Scale Dark Acoustic Oscillations from Cosmology,''
Phys. Rev. D \textbf{89}, no.6, 063517 (2014)
doi:10.1103/PhysRevD.89.063517
[arXiv:1310.3278 [astro-ph.CO]].

\bibitem{Buckley:2014hja}
M.~R.~Buckley, J.~Zavala, F.~Y.~Cyr-Racine, K.~Sigurdson and M.~Vogelsberger,
``Scattering, Damping, and Acoustic Oscillations: Simulating the Structure of Dark Matter Halos with Relativistic Force Carriers,''
Phys. Rev. D \textbf{90}, no.4, 043524 (2014)
doi:10.1103/PhysRevD.90.043524
[arXiv:1405.2075 [astro-ph.CO]].





\bibitem{Kahlhoefer:2017umn}
F.~Kahlhoefer, K.~Schmidt-Hoberg and S.~Wild,
``Dark matter self-interactions from a general spin-0 mediator,''
JCAP \textbf{08}, 003 (2017)
doi:10.1088/1475-7516/2017/08/003
[arXiv:1704.02149 [hep-ph]].

\bibitem{Khrapak:2003kjw}
S.~A.~Khrapak, A.~V.~Ivlev, G.~E.~Morfill and S.~K.~Zhdanov,
``Scattering in the Attractive Yukawa Potential in the Limit of Strong Interaction,''
Phys. Rev. Lett. \textbf{90}, no.22, 225002 (2003)
doi:10.1103/PhysRevLett.90.225002

\bibitem{Cyr-Racine:2015ihg}
F.~Y.~Cyr-Racine, K.~Sigurdson, J.~Zavala, T.~Bringmann, M.~Vogelsberger and C.~Pfrommer,
``ETHOS\textemdash{}an effective theory of structure formation: From dark particle physics to the matter distribution of the Universe,''
Phys. Rev. D \textbf{93}, no.12, 123527 (2016)
doi:10.1103/PhysRevD.93.123527
[arXiv:1512.05344 [astro-ph.CO]].

\bibitem{Colquhoun:2020adl}
B.~Colquhoun, S.~Heeba, F.~Kahlhoefer, L.~Sagunski and S.~Tulin,
``Semiclassical regime for dark matter self-interactions,''
Phys. Rev. D \textbf{103}, no.3, 035006 (2021)
doi:10.1103/PhysRevD.103.035006
[arXiv:2011.04679 [hep-ph]].

\bibitem{GasD1965}
W. G. Vincenti and C. H. Kruger, \emph{ Introduction to Physical Gas Dynamics} (Krieger, Malabar, FL, 1965).




\bibitem{Witten:1984rs}
E.~Witten,
``Cosmic Separation of Phases,''
Phys. Rev. D \textbf{30}, 272-285 (1984)
doi:10.1103/PhysRevD.30.272





\bibitem{Jaeckel:2016jlh}
J.~Jaeckel, V.~V.~Khoze and M.~Spannowsky,
``Hearing the signal of dark sectors with gravitational wave detectors,''
Phys. Rev. D \textbf{94}, no.10, 103519 (2016)
doi:10.1103/PhysRevD.94.103519
[arXiv:1602.03901 [hep-ph]].

\bibitem{Schwaller:2015tja}
P.~Schwaller,
``Gravitational Waves from a Dark Phase Transition,''
Phys. Rev. Lett. \textbf{115}, no.18, 181101 (2015)
doi:10.1103/PhysRevLett.115.181101
[arXiv:1504.07263 [hep-ph]].

\bibitem{Soni:2016yes}
A.~Soni and Y.~Zhang,
``Gravitational Waves From SU(N) Glueball Dark Matter,''
Phys. Lett. B \textbf{771}, 379-384 (2017)
doi:10.1016/j.physletb.2017.05.077
[arXiv:1610.06931 [hep-ph]].


\bibitem{Addazi:2017gpt}
A.~Addazi and A.~Marciano,
``Gravitational waves from dark first order phase transitions and dark photons,''
Chin. Phys. C \textbf{42}, no.2, 023107 (2018)
doi:10.1088/1674-1137/42/2/023107
[arXiv:1703.03248 [hep-ph]].

\bibitem{Tsumura:2017knk}
K.~Tsumura, M.~Yamada and Y.~Yamaguchi,
``Gravitational wave from dark sector with dark pion,''
JCAP \textbf{07}, 044 (2017)
doi:10.1088/1475-7516/2017/07/044
[arXiv:1704.00219 [hep-ph]].

\bibitem{Huang:2017rzf}
F.~P.~Huang and J.~H.~Yu,
``Exploring inert dark matter blind spots with gravitational wave signatures,''
Phys. Rev. D \textbf{98}, no.9, 095022 (2018)
doi:10.1103/PhysRevD.98.095022
[arXiv:1704.04201 [hep-ph]].

\bibitem{Hashino:2018zsi}
K.~Hashino, M.~Kakizaki, S.~Kanemura, P.~Ko and T.~Matsui,
``Gravitational waves from first order electroweak phase transition in models with the U(1)$_{X}$ gauge symmetry,''
JHEP \textbf{06}, 088 (2018)
doi:10.1007/JHEP06(2018)088
[arXiv:1802.02947 [hep-ph]].

\bibitem{Bai:2018dxf}
Y.~Bai, A.~J.~Long and S.~Lu,
``Dark Quark Nuggets,''
Phys. Rev. D \textbf{99}, no.5, 055047 (2019)
doi:10.1103/PhysRevD.99.055047
[arXiv:1810.04360 [hep-ph]].

\bibitem{Breitbach:2018ddu}
M.~Breitbach, J.~Kopp, E.~Madge, T.~Opferkuch and P.~Schwaller,
``Dark, Cold, and Noisy: Constraining Secluded Hidden Sectors with Gravitational Waves,''
JCAP \textbf{07}, 007 (2019)
doi:10.1088/1475-7516/2019/07/007
[arXiv:1811.11175 [hep-ph]].

\bibitem{Fairbairn:2019xog}
M.~Fairbairn, E.~Hardy and A.~Wickens,
``Hearing without seeing: gravitational waves from hot and cold hidden sectors,''
JHEP \textbf{07}, 044 (2019)
doi:10.1007/JHEP07(2019)044
[arXiv:1901.11038 [hep-ph]].

\bibitem{Addazi:2020zcj}
A.~Addazi, Y.~F.~Cai, Q.~Gan, A.~Marciano and K.~Zeng,
``NANOGrav results and dark first order phase transitions,''
Sci. China Phys. Mech. Astron. \textbf{64}, no.9, 290411 (2021)
doi:10.1007/s11433-021-1724-6
[arXiv:2009.10327 [hep-ph]].

\bibitem{Ratzinger:2020koh}
W.~Ratzinger and P.~Schwaller,
``Whispers from the dark side: Confronting light new physics with NANOGrav data,''
SciPost Phys. \textbf{10}, no.2, 047 (2021)
doi:10.21468/SciPostPhys.10.2.047
[arXiv:2009.11875 [astro-ph.CO]].

\bibitem{Ghosh:2020ipy}
T.~Ghosh, H.~K.~Guo, T.~Han and H.~Liu,
``Electroweak phase transition with an SU(2) dark sector,''
JHEP \textbf{07}, 045 (2021)
doi:10.1007/JHEP07(2021)045
[arXiv:2012.09758 [hep-ph]].

\bibitem{Dent:2022bcd}
J.~B.~Dent, B.~Dutta, S.~Ghosh, J.~Kumar and J.~Runburg,
``Sensitivity to dark sector scales from gravitational wave signatures,''
JHEP \textbf{08}, 300 (2022)
doi:10.1007/JHEP08(2022)300
[arXiv:2203.11736 [hep-ph]].

\bibitem{Wang:2022lxn}
W.~Wang, K.~P.~Xie, W.~L.~Xu and J.~M.~Yang,
``Cosmological phase transitions, gravitational waves and self-interacting dark matter in the singlet extension of MSSM,''
Eur. Phys. J. C \textbf{82}, no.12, 1120 (2022)
doi:10.1140/epjc/s10052-022-11077-3
[arXiv:2204.01928 [hep-ph]].

\bibitem{Wang:2022akn}
W.~Wang, W.~L.~Xu and J.~M.~Yang,
``A hidden self-interacting dark matter sector with first order cosmological phase transition and gravitational wave,''
[arXiv:2209.11408 [hep-ph]].






\bibitem{Wainwright:2011qy}
C.~Wainwright, S.~Profumo and M.~J.~Ramsey-Musolf,
``Gravity Waves from a Cosmological Phase Transition: Gauge Artifacts and Daisy Resummations,''
Phys. Rev. D \textbf{84}, 023521 (2011)
doi:10.1103/PhysRevD.84.023521
[arXiv:1104.5487 [hep-ph]].

\bibitem{Chiang:2017zbz}
C.~W.~Chiang and E.~Senaha,
``On gauge dependence of gravitational waves from a first-order phase transition in classical scale-invariant $U(1)'$ models,''
Phys. Lett. B \textbf{774}, 489-493 (2017)
doi:10.1016/j.physletb.2017.09.064
[arXiv:1707.06765 [hep-ph]].


\bibitem{Karjalainen:1996wx}
M.~Karjalainen, M.~Laine and J.~Peisa,
``The Order of the phase transition in 3-d U(1) + Higgs theory,''
Nucl. Phys. B Proc. Suppl. \textbf{53}, 475-480 (1997)
doi:10.1016/S0920-5632(96)00692-5
[arXiv:hep-lat/9608006 [hep-lat]].

\bibitem{Dimopoulos:1997cz}
P.~Dimopoulos, K.~Farakos and G.~Koutsoumbas,
``Three-dimensional lattice U(1) gauge Higgs model at low m(H),''
Eur. Phys. J. C \textbf{1}, 711-719 (1998)
doi:10.1007/s100520050116
[arXiv:hep-lat/9703004 [hep-lat]].











\bibitem{Coleman:1973jx} 
  S.~R.~Coleman and E.~J.~Weinberg,
  ``Radiative Corrections as the Origin of Spontaneous Symmetry Breaking,''
  Phys.\ Rev.\ D {\bf 7}, 1888 (1973).
  doi:10.1103/PhysRevD.7.1888

\bibitem{thermal_correction1}
  L.~Dolan and R.~Jackiw,
  ``Symmetry Behavior at Finite Temperature,''
  Phys.\ Rev.\ D {\bf 9}, 3320 (1974).
  doi:10.1103/PhysRevD.9.3320
 \bibitem{thermal_correction2}
  M.~Quiros,
  ``Finite temperature field theory and phase transitions,''
  hep-ph/9901312.


\bibitem{Arnold:1992rz} 
  P.~B.~Arnold and O.~Espinosa,
  ``The Effective potential and first order phase transitions: Beyond leading-order,''
  Phys.\ Rev.\ D {\bf 47}, 3546 (1993)
  Erratum: [Phys.\ Rev.\ D {\bf 50}, 6662 (1994)]
  doi:10.1103/physrevd.50.6662.2, 10.1103/PhysRevD.47.3546
  [hep-ph/9212235].


\bibitem{Delaunay:2007wb}
C.~Delaunay, C.~Grojean and J.~D.~Wells,
``Dynamics of Non-renormalizable Electroweak Symmetry Breaking,''
JHEP \textbf{04}, 029 (2008)
doi:10.1088/1126-6708/2008/04/029
[arXiv:0711.2511 [hep-ph]].

\bibitem{Anderson:1991zb}
G.~W.~Anderson and L.~J.~Hall,
``The Electroweak phase transition and baryogenesis,''
Phys. Rev. D \textbf{45}, 2685-2698 (1992)
doi:10.1103/PhysRevD.45.2685







\bibitem{Coleman:1977py}
S.~R.~Coleman,
``The Fate of the False Vacuum. 1. Semiclassical Theory,''
Phys. Rev. D \textbf{15}, 2929-2936 (1977)
[erratum: Phys. Rev. D \textbf{16}, 1248 (1977)]
doi:10.1103/PhysRevD.16.1248

\bibitem{Callan:1977pt}
C.~G.~Callan, Jr. and S.~R.~Coleman,
``The Fate of the False Vacuum. 2. First Quantum Corrections,''
Phys. Rev. D \textbf{16}, 1762-1768 (1977)
doi:10.1103/PhysRevD.16.1762

\bibitem{Linde:1981zj}
A.~D.~Linde,
``Decay of the False Vacuum at Finite Temperature,''
Nucl. Phys. B \textbf{216}, 421 (1983)
[erratum: Nucl. Phys. B \textbf{223}, 544 (1983)]
doi:10.1016/0550-3213(83)90072-X

\bibitem{Apreda:2001us}
R.~Apreda, M.~Maggiore, A.~Nicolis and A.~Riotto,
``Gravitational waves from electroweak phase transitions,''
Nucl. Phys. B \textbf{631}, 342-368 (2002)
doi:10.1016/S0550-3213(02)00264-X
[arXiv:gr-qc/0107033 [gr-qc]].

\bibitem{Wainwright:2011kj}
C.~L.~Wainwright,
``CosmoTransitions: Computing Cosmological Phase Transition Temperatures and Bubble Profiles with Multiple Fields,''
Comput. Phys. Commun. \textbf{183}, 2006-2013 (2012)
doi:10.1016/j.cpc.2012.04.004
[arXiv:1109.4189 [hep-ph]].








\bibitem{Kosowsky:1991ua}
A.~Kosowsky, M.~S.~Turner and R.~Watkins,
``Gravitational radiation from colliding vacuum bubbles,''
Phys. Rev. D \textbf{45}, 4514-4535 (1992)
doi:10.1103/PhysRevD.45.4514

\bibitem{Kosowsky:1992rz}
A.~Kosowsky, M.~S.~Turner and R.~Watkins,
``Gravitational waves from first order cosmological phase transitions,''
Phys. Rev. Lett. \textbf{69}, 2026-2029 (1992)
doi:10.1103/PhysRevLett.69.2026

\bibitem{Kosowsky:1992vn}
A.~Kosowsky and M.~S.~Turner,
``Gravitational radiation from colliding vacuum bubbles: envelope approximation to many bubble collisions,''
Phys. Rev. D \textbf{47}, 4372-4391 (1993)
doi:10.1103/PhysRevD.47.4372
[arXiv:astro-ph/9211004 [astro-ph]].

\bibitem{Kamionkowski:1993fg}
M.~Kamionkowski, A.~Kosowsky and M.~S.~Turner,
``Gravitational radiation from first order phase transitions,''
Phys. Rev. D \textbf{49}, 2837-2851 (1994)
doi:10.1103/PhysRevD.49.2837
[arXiv:astro-ph/9310044 [astro-ph]].

\bibitem{Caprini:2007xq}
C.~Caprini, R.~Durrer and G.~Servant,
``Gravitational wave generation from bubble collisions in first-order phase transitions: An analytic approach,''
Phys. Rev. D \textbf{77}, 124015 (2008)
doi:10.1103/PhysRevD.77.124015
[arXiv:0711.2593 [astro-ph]].

\bibitem{Huber:2008hg}
S.~J.~Huber and T.~Konstandin,
``Gravitational Wave Production by Collisions: More Bubbles,''
JCAP \textbf{09}, 022 (2008)
doi:10.1088/1475-7516/2008/09/022
[arXiv:0806.1828 [hep-ph]].







\bibitem{Hindmarsh:2013xza}
M.~Hindmarsh, S.~J.~Huber, K.~Rummukainen and D.~J.~Weir,
``Gravitational waves from the sound of a first order phase transition,''
Phys. Rev. Lett. \textbf{112}, 041301 (2014)
doi:10.1103/PhysRevLett.112.041301
[arXiv:1304.2433 [hep-ph]].

\bibitem{Giblin:2013kea}
J.~T.~Giblin, Jr. and J.~B.~Mertens,
``Vacuum Bubbles in the Presence of a Relativistic Fluid,''
JHEP \textbf{12}, 042 (2013)
doi:10.1007/JHEP12(2013)042
[arXiv:1310.2948 [hep-th]].

\bibitem{Giblin:2014qia}
J.~T.~Giblin and J.~B.~Mertens,
``Gravitional radiation from first-order phase transitions in the presence of a fluid,''
Phys. Rev. D \textbf{90}, no.2, 023532 (2014)
doi:10.1103/PhysRevD.90.023532
[arXiv:1405.4005 [astro-ph.CO]].

\bibitem{Hindmarsh:2015qta}
M.~Hindmarsh, S.~J.~Huber, K.~Rummukainen and D.~J.~Weir,
``Numerical simulations of acoustically generated gravitational waves at a first order phase transition,''
Phys. Rev. D \textbf{92}, no.12, 123009 (2015)
doi:10.1103/PhysRevD.92.123009
[arXiv:1504.03291 [astro-ph.CO]].





\bibitem{Caprini:2006jb}
C.~Caprini and R.~Durrer,
``Gravitational waves from stochastic relativistic sources: Primordial turbulence and magnetic fields,''
Phys. Rev. D \textbf{74}, 063521 (2006)
doi:10.1103/PhysRevD.74.063521
[arXiv:astro-ph/0603476 [astro-ph]].

\bibitem{Kahniashvili:2008pf}
T.~Kahniashvili, A.~Kosowsky, G.~Gogoberidze and Y.~Maravin,
``Detectability of Gravitational Waves from Phase Transitions,''
Phys. Rev. D \textbf{78}, 043003 (2008)
doi:10.1103/PhysRevD.78.043003
[arXiv:0806.0293 [astro-ph]].

\bibitem{Kahniashvili:2008pe}
T.~Kahniashvili, L.~Campanelli, G.~Gogoberidze, Y.~Maravin and B.~Ratra,
``Gravitational Radiation from Primordial Helical Inverse Cascade MHD Turbulence,''
Phys. Rev. D \textbf{78}, 123006 (2008)
[erratum: Phys. Rev. D \textbf{79}, 109901 (2009)]
doi:10.1103/PhysRevD.78.123006
[arXiv:0809.1899 [astro-ph]].

\bibitem{Kahniashvili:2009mf}
T.~Kahniashvili, L.~Kisslinger and T.~Stevens,
``Gravitational Radiation Generated by Magnetic Fields in Cosmological Phase Transitions,''
Phys. Rev. D \textbf{81}, 023004 (2010)
doi:10.1103/PhysRevD.81.023004
[arXiv:0905.0643 [astro-ph.CO]].

\bibitem{Caprini:2009yp}
C.~Caprini, R.~Durrer and G.~Servant,
``The stochastic gravitational wave background from turbulence and magnetic fields generated by a first-order phase transition,''
JCAP \textbf{12}, 024 (2009)
doi:10.1088/1475-7516/2009/12/024
[arXiv:0909.0622 [astro-ph.CO]].


\bibitem{Kisslinger:2015hua}
L.~Kisslinger and T.~Kahniashvili,
``Polarized Gravitational Waves from Cosmological Phase Transitions,''
Phys. Rev. D \textbf{92}, no.4, 043006 (2015)
doi:10.1103/PhysRevD.92.043006
[arXiv:1505.03680 [astro-ph.CO]].







\bibitem{Ellis:2018mja}
J.~Ellis, M.~Lewicki and J.~M.~No,
``On the Maximal Strength of a First-Order Electroweak Phase Transition and its Gravitational Wave Signal,''
JCAP \textbf{04}, 003 (2019)
doi:10.1088/1475-7516/2019/04/003
[arXiv:1809.08242 [hep-ph]].

\bibitem{Ellis:2020awk}
J.~Ellis, M.~Lewicki and J.~M.~No,
``Gravitational waves from first-order cosmological phase transitions: lifetime of the sound wave source,''
JCAP \textbf{07}, 050 (2020)
doi:10.1088/1475-7516/2020/07/050
[arXiv:2003.07360 [hep-ph]].

\bibitem{Wang:2020jrd}
X.~Wang, F.~P.~Huang and X.~Zhang,
``Phase transition dynamics and gravitational wave spectra of strong first-order phase transition in supercooled universe,''
JCAP \textbf{05}, 045 (2020)
doi:10.1088/1475-7516/2020/05/045
[arXiv:2003.08892 [hep-ph]].





\bibitem{Espinosa:2010hh}
J.~R.~Espinosa, T.~Konstandin, J.~M.~No and G.~Servant,
``Energy Budget of Cosmological First-order Phase Transitions,''
JCAP \textbf{06}, 028 (2010)
doi:10.1088/1475-7516/2010/06/028
[arXiv:1004.4187 [hep-ph]].


\bibitem{Schmitz:2020syl}
K.~Schmitz,
``New Sensitivity Curves for Gravitational-Wave Signals from Cosmological Phase Transitions,''
JHEP \textbf{01}, 097 (2021)
doi:10.1007/JHEP01(2021)097
[arXiv:2002.04615 [hep-ph]].


\bibitem{Moore:1995si}
G.~D.~Moore and T.~Prokopec,
``How fast can the wall move? A Study of the electroweak phase transition dynamics,''
Phys. Rev. D \textbf{52}, 7182-7204 (1995)
doi:10.1103/PhysRevD.52.7182
[arXiv:hep-ph/9506475 [hep-ph]].

\bibitem{Megevand:2009gh}
A.~Megevand and A.~D.~Sanchez,
``Velocity of electroweak bubble walls,''
Nucl. Phys. B \textbf{825}, 151-176 (2010)
doi:10.1016/j.nuclphysb.2009.09.019
[arXiv:0908.3663 [hep-ph]].

\bibitem{Huber:2013kj}
S.~J.~Huber and M.~Sopena,
``An efficient approach to electroweak bubble velocities,''
[arXiv:1302.1044 [hep-ph]].

\bibitem{Konstandin:2014zta}
T.~Konstandin, G.~Nardini and I.~Rues,
``From Boltzmann equations to steady wall velocities,''
JCAP \textbf{09}, 028 (2014)
doi:10.1088/1475-7516/2014/09/028
[arXiv:1407.3132 [hep-ph]].

\bibitem{Dorsch:2018pat}
G.~C.~Dorsch, S.~J.~Huber and T.~Konstandin,
``Bubble wall velocities in the Standard Model and beyond,''
JCAP \textbf{12}, 034 (2018)
doi:10.1088/1475-7516/2018/12/034
[arXiv:1809.04907 [hep-ph]].

\bibitem{Laurent:2022jrs}
B.~Laurent and J.~M.~Cline,
``First principles determination of bubble wall velocity,''
Phys. Rev. D \textbf{106}, no.2, 023501 (2022)
doi:10.1103/PhysRevD.106.023501
[arXiv:2204.13120 [hep-ph]].

\bibitem{Wang:2020zlf}
X.~Wang, F.~P.~Huang and X.~Zhang,
``Bubble wall velocity beyond leading-log approximation in electroweak phase transition,''
[arXiv:2011.12903 [hep-ph]].






\bibitem{Caprini:2015zlo}
C.~Caprini, M.~Hindmarsh, S.~Huber, T.~Konstandin, J.~Kozaczuk, G.~Nardini, J.~M.~No, A.~Petiteau, P.~Schwaller and G.~Servant, \textit{et al.}
``Science with the space-based interferometer eLISA. II: Gravitational waves from cosmological phase transitions,''
JCAP \textbf{04}, 001 (2016)
doi:10.1088/1475-7516/2016/04/001
[arXiv:1512.06239 [astro-ph.CO]].

\bibitem{Caprini:2019egz}
C.~Caprini, M.~Chala, G.~C.~Dorsch, M.~Hindmarsh, S.~J.~Huber, T.~Konstandin, J.~Kozaczuk, G.~Nardini, J.~M.~No and K.~Rummukainen, \textit{et al.}
``Detecting gravitational waves from cosmological phase transitions with LISA: an update,''
JCAP \textbf{03}, 024 (2020)
doi:10.1088/1475-7516/2020/03/024
[arXiv:1910.13125 [astro-ph.CO]].

\bibitem{Huber:2008hg}
S.~J.~Huber and T.~Konstandin,
``Gravitational Wave Production by Collisions: More Bubbles,''
JCAP \textbf{09}, 022 (2008)
doi:10.1088/1475-7516/2008/09/022
[arXiv:0806.1828 [hep-ph]].

\bibitem{Hindmarsh:2015qta}
M.~Hindmarsh, S.~J.~Huber, K.~Rummukainen and D.~J.~Weir,
``Numerical simulations of acoustically generated gravitational waves at a first order phase transition,''
Phys. Rev. D \textbf{92}, no.12, 123009 (2015)
doi:10.1103/PhysRevD.92.123009
[arXiv:1504.03291 [astro-ph.CO]].

\bibitem{Caprini:2009yp}
C.~Caprini, R.~Durrer and G.~Servant,
``The stochastic gravitational wave background from turbulence and magnetic fields generated by a first-order phase transition,''
JCAP \textbf{12}, 024 (2009)
doi:10.1088/1475-7516/2009/12/024
[arXiv:0909.0622 [astro-ph.CO]].







\bibitem{Janssen:2014dka}
G.~Janssen, G.~Hobbs, M.~McLaughlin, C.~Bassa, A.~T.~Deller, M.~Kramer, K.~Lee, C.~Mingarelli, P.~Rosado and S.~Sanidas, \textit{et al.}
``Gravitational wave astronomy with the SKA,''
PoS \textbf{AASKA14}, 037 (2015)
doi:10.22323/1.215.0037
[arXiv:1501.00127 [astro-ph.IM]].

\bibitem{LISA:2017pwj}
P.~Amaro-Seoane \textit{et al.} [LISA],
``Laser Interferometer Space Antenna,''
[arXiv:1702.00786 [astro-ph.IM]].



\end{thebibliography}

\end{document}